\documentclass{article}
\usepackage{lineno}
\usepackage{ae,aecompl}
\usepackage[utf8]{inputenc}
\usepackage[T1]{fontenc}
\usepackage[english]{babel}
\usepackage[style=numeric-comp,backend=biber,natbib=true,sorting=none]{biblatex}
\usepackage{csquotes}
\usepackage{amssymb, amsmath, amsthm, amsfonts}
\usepackage{mathtools}
\usepackage{siunitx}
\usepackage{caption}
\usepackage{multirow}
\usepackage{derivative}
\usepackage{physics}
\usepackage[hidelinks]{hyperref}
\usepackage{enumerate, cleveref}
\usepackage{tabstackengine}
\usepackage{stackengine}
\usepackage{algorithm, algpseudocode}
\usepackage{supertabular, makecell, booktabs,wrapfig, longtable}
\usepackage{orcidlink}
\usepackage{textcomp}
\usepackage{subfigure}
\usepackage{placeins, relsize, fix-cm}
\usepackage{todonotes}
\usepackage{tikz}
\usetikzlibrary{shapes.geometric, arrows, patterns, patterns.meta}

\usepackage{geometry}
\usepackage{authblk}

\newtheorem{theorem}{Theorem}[section]
\theoremstyle{remark}
\newtheorem{remark}[theorem]{Remark}
\newtheorem*{remark*}{Remark}

\numberwithin{equation}{section}

\patchcmd{\subequations}  
{\theparentequation\alph{equation}}  
{\theparentequation.\alph{equation}}  
{}{}  

\providecommand{\nphase}{n_P} 
\providecommand{\ncomp}{n_C} 
\providecommand{\extfrac}{\chi}  
\providecommand{\varX}{X}  
\providecommand{\varY}{Y}  
\providecommand{\absperm}{\mathbf{K}}  
\providecommand{\fugCoeff}{\varphi}  
\providecommand{\LagMultipier}{\boldsymbol{\lambda}}  
\providecommand{\specGibbs}{g}
\providecommand{\specEnthalpy}{h}

\providecommand{\sysF}{\mathrm{F}}  
\providecommand{\jac}{\mathrm{J}}  
\providecommand{\resph}{\sysF_{p\specEnthalpy}}
\providecommand{\resft}{\sysF_{ft}}
\providecommand{\jacof}[1]{\ensuremath{\mathbf{J}_{\!#1}}}

\newcommand{\cotwo}{CO$_2$\,}

\tikzstyle{arrow} = [thick,->,>=stealth]
\tikzstyle{process} = [rectangle, minimum width=1.5cm, minimum height=0.5cm, text width=1.5cm, text centered, draw=black]
\tikzstyle{decision} = [diamond, minimum width=0.1cm, minimum height=0.1cm, aspect=2, text centered, draw=black, fill=black!5]
\tikzstyle{optprocess} = [rectangle, minimum width=1.5cm, minimum height=0.5cm, text width=1.5cm, text centered, draw=black, dashed]
\tikzstyle{every node}=[font=\tiny]

\makeatletter 
\def\WFfill{\par 
    \ifx\parshape\WF@fudgeparshape 
    \nobreak 
    \ifnum\c@WF@wrappedlines>\@ne 
    \advance\c@WF@wrappedlines\m@ne 
    \vskip\c@WF@wrappedlines\baselineskip 
    \global\c@WF@wrappedlines\z@ 
    \fi 
    \allowbreak 
    \WF@finale 
    \fi 
} 
\makeatother

\providecommand{\keywords}[1]
{
	\small	
	\textit{Keywords---} #1
}

\providecommand{\correspondant}[1]
{
	\small	
	\textit{Corresponding author:} #1
}

\title{Persistent-variable thermal compositional simulation of multiphase flow with phase separation in porous media}
\date{\today}
\author[1]{Veljko Lipovac \orcidlink{0000-0002-9442-0879}}
\author[1,2]{Omar Duran \orcidlink{0000-0002-0343-9890}}
\author[1]{Eirik Keilegavlen \orcidlink{0000-0002-0333-9507}}
\author[1]{Inga Berre \orcidlink{0000-0002-0212-7959}}

\affil[1]{Center for Modeling of Coupled Subsurface Dynamics \\ Department of Mathematics, University of Bergen, Bergen, Norway}

\affil[2]{Department of Energy Science and Engineering (ESE) \\ Stanford Doerr School of Sustainability, Stanford University, USA}

\begin{document}

\maketitle

\begin{abstract}
    Thermal compositional multiphase flow in porous media with phase transitions involves complex nonlinear interactions among fluid flow, component transport, and thermodynamic phase equilibrium, posing significant computational challenges.
This paper presents a novel persistent-variable formulation for thermal compositional flow using enthalpy to formulate the energy balance and the local equilibrium problem.
Equilibrium conditions are derived from a thermodynamically consistent minimization problem using a persistent set of variables, allowing for seamless integration of equilibrium calculations into a fully coupled flow and transport model.
This formulation does not require phase stability tests, fixes the number of possible phases independent of their physical presence, and provides a continuous and full mathematical description of the multiphysics system.
The enthalpy-based formulation additionally addresses challenging non-isothermal scenarios.
To tackle the nonlinearities arising from phase transitions, we embed a local solver for the thermodynamic subproblem within a global Newton solver for the fully implicit system.
The local solver exploits the locality of the thermodynamic subproblem and allows for an efficient parallelized solution.
It also leverages the modularity of the persistent-variable formulation and uses both isothermal and isenthalpic equilibrium conditions locally.
We demonstrate the capability of our approach to simulate complex high-enthalpy systems, including narrow-boiling phenomena.
The impact of the embedded local solver is analyzed through numerical experiments, demonstrating a reduction in global nonlinear iterations of up to 23 \% with increased use of the local solver.
The number of local iterations is controlled with a local solver tolerance and no significant impact on the global iteration number was observed for local residual tolerances as high as $\num{1e-3}$.
The persistent-variable approach using enthalpy and the flexibility and scalability of the embedded local solver advance the usage of equilibrium calculations in multiphase flow simulations and are suitable for high-enthalpy applications.
\end{abstract}

\keywords{porous media,
    multiphase flow,
    compositional simulation,
    phase separation,
    nonlinear solver,
    persistent-variable formulation,
    high-enthalpy
}

\correspondant{Veljko Lipovac \texttt{veljko.lipovac@uib.no}}

\section{Introduction}
Simulations of thermal compositional flow with phase change in porous media support critical applications ranging from geothermal energy extraction to hydrocarbon reservoir management and subsurface carbon sequestration.
These systems are characterized by complex interactions between fluid flow, component transport, and thermodynamic phase behavior, often under non-isothermal conditions.
Accurate simulation of such processes requires a robust framework that couples the governing equations of mass and energy conservation with thermodynamically consistent phase separation calculations, as well as a solution strategy able to tackle the challenges introduced by individual physics.

One central challenge is the fact that the number of phases present at equilibrium is an unknown of the problem.
That is, the number of degrees of freedom is per se unknown, which introduces various technical and mathematical difficulties in the flow and transport model, as well as its implementation.
Modern thermodynamics and phase separation calculations (\textit{flash} algorithms) entail two essential steps as formalized by \citet{michelsen1982-1,michelsen1982-2}, the phase split calculations and the phase stability tests.
While the phase split calculations are an application of the principles of minimal energy or maximal entropy for some target fluid state \cite{michelsen1999}, stability tests are performed to check whether adding or removing a phase and its contribution to the minimized function would result in a lower value.
Stability tests are required for all target state definitions \cite{nichita2024}.
Although methodologically challenging, this approach is the most widely used, including for challenging narrow-boiling fluids such as water, where changes in enthalpy can be large for small changes in temperature \cite{zhu2014,zhu2016}.

An alternative formulation includes a persistent set of variables. Possible phases and respective degrees of freedom can be anticipated and the degrees of freedom are persistently included in the mathematical formulation.
\citet{gupta1990} present a formulation of the respective minimization problem with inequality constraints, which allow degrees of freedom associated with vanished phases to take the value zero.
\citet{gharbia2021} prove a connection between the inequality-constrained optimization problem and stability tests.
The authors of this paper previously extended the persistent-variable approach to equilibrium conditions involving isenthalpic and isochoric specifications \cite{lipovac2024}.

Formulations for thermal compositional flow in porous media have also evolved over time to account for the increasing complexity \cite{lake1989,chen2000}.
\citet{voskov2012} provide an extensive overview of the possibilities of choosing variables and equations for flow and transport, intended for applications with phase change and thermodynamically consistent fluid models.
In essence, the starting point for any balance equation is the balance equation of mass or energy for a single component/species in a specific phase.
We model a set of components, while we have only an idea of the set of possible phases since they may physically not be present at all times.
The modeling process is continued by one of the following two approaches:
On the one hand, a summation over the set of components in a phase results in saturation-based balance equations per phase.
These formulations are well understood for immiscible-flow and volume-based formulations, mainly used in multiphase models without phase change \cite{lake1989,chen2000}.
On the other hand, a summation over the phases results in an \textit{overall composition} formulation of balance equations per fluid component.
The sum is formally performed over all phases without taking phase disappearance into account.
We call the changing set of present phases the current \textit{phase context}.
This approach is widely used in compositional and reactive flow models because it introduces the notion of total mass of a component.
It is a necessary input quantity for both phase separation calculations and chemical reactions \cite{kala2020, moortgat2011}.

We observe a general trend to use components/species as a basis to formulate the transport problem, instead of phases.
This undoubtedly connects to the fact that the phase context is unknown due to the nature of the local problem.
However, compared to the model formulation, the application of local equilibrium calculations as part of a global solution procedure to resolve the phase context has not been as extensively studied.
\citet{class2002} present a thermal compositional flow model with a switching-variable approach, where the set of primary variables is changed depending on the phase context resulting from the phase separation calculations.
This switch is carried out in every iteration of the global nonlinear solver.
\citet{flemisch2011} solve the flow and transport problem based on concentrations and perform phase separation calculations once the transport problem converges.
\citet{novikov2025} and \citet{voskov2017} present a solution in which they express the entire flow and transport problem as an interpolation in the fluid state space, exemplarily described by pressure, temperature, and mass.
Once the simulation moves in a previously uncovered area, the phase separation calculations are performed and stored for future usage.
\citet{wang2020} present the same approach using enthalpy instead of temperature, to simulate challenging non-isothermal cases.
\citet{Faigle2015} point out that it is possible to outsource the entire thermodynamic subproblem and focus on other pressing challenges such as operator splitting and grid adaptation.
However, feeding back critical information, such as derivatives of fluid properties with respect to flow and transport variables, is unclear with this approach.

Despite significant research contributions to modeling of thermal compositional flow over the last decades, it is a matter of active research to find a formulation that captures the whole coupled dynamics in a mathematically and thermodynamically consistent way.
We point out two things:
First, due to the nature of the thermodynamic problem and the changing phase context, most of the approaches so far introduce a degree of discontinuity into the solution procedure.
This is independent of whether it is a switching-variable approach or an overall composition formulation with terms appearing and disappearing in the model equations.
If using phase stability tests to change the phase context, the underlying physical principle, the Gibbs tangent plane criterion \cite{michelsen1982-1,gharbia2021}, does not enter the analytical formulation of the global system and consequently its residual.
Hence, a changing phase context will appear as a discontinuity during the global nonlinear iterations, with the risk of deteriorating convergence.
Although a first coupling of the persistent-variable formulation with flow and transport was achieved by \citet{lauser2011}, it was done with a relatively simple fluid model and only for a temperature-based formulation, which is not suitable for high-enthalpy scenarios.
And second, there is no clear answer to when, how and if at all to use a local solver to perform the phase separation calculations and how to incorporate it into the global nonlinear solver.
With nonlinearities and increasing complexity of fluid models or equations of state, this must be taken into considerations on robustness and efficiency, besides other known computational challenges relating to subsurface modeling.

To address above issues, we present a persistent-variable formulation of thermal compositional flow based on enthalpy, which includes a local, isenthalpic equilibrium problem without the need for phase stability tests.
Using both fluid enthalpy and temperature in the accumulation term and diffusive flux of the energy balance equation respectively, we can show that the isenthalpic local equilibrium problem is a closed subsystem of the whole multi-physics model.
By consequence of the persistent formulation of the isenthalpic problem, the isothermal equilibrium system is yet another closed and persistent subsystem.
The phase context is fixed, and phase appearance and disappearance are handled using inequality constraints, which leads to a semi-smooth but continuous, and complete analytical representation of the coupled model.

We also present a detailed nonlinear solution strategy, which embeds a local solver for the equilibrium subproblem within a global Newton solver for the fully implicit system.
The computational cost of phase separation calculations is mitigated by exploiting the locality of the subproblem, parallelization, and its optimization-based formulation. 
The modularity of the persistent-variable formulation of the equilibrium problem allows us to conduct phase separation calculations locally under different target conditions, isothermal and isenthalpic ones.
We successfully applied this approach to a complex high-enthalpy system, including narrow-boiling conditions, and a challenging equation-of-state-based fluid model.
And finally, we investigated how the frequency of applying the local solver affects the nonlinear global iterations, and to what precision the local problem must be solved.

\Cref{subsection:flow-transport} presents an overall composition formulation of flow and transport of mass and energy.
\Cref{subsection:phase-equilibrium,subsection:model-summary} introduce the local equilibrium from the first principles of thermodynamics, and couple the resulting local algebraic equations with the partial differential equations.
\Cref{section:numerical-method} outlines the nonlinear solution strategy, involving a global Newton method and a local solver to resolve phase separation.
To test the presented model and solution strategy, we manufactured an example and simulated the injection of a cold, two-component fluid mixture into a hot domain with narrow-boiling water.
\Cref{section:results} presents the analysis of the efficiency of the proposed solution strategy, switching between temperature- and enthalpy-based equilibrium conditions and investigating the parameterization of the local solver.

\section{Mathematical model}\label{section:mathematical-model}
In this section, we present the mathematical model for thermal flow and transport in porous media in strong form, as well as the model for phase separation (equilibrium) as a subsystem of local equations.
We utilize an overall composition formulation for the transport problem, introducing the fraction of mass associated with a fluid component as a transport variable, which is also required for the local equilibration.
The phase equilibrium problem is derived from a thermodynamically consistent minimization approach, utilizing the formulation with persistent sets of variables and equations.
Furthermore, we introduce an isenthalpic formulation involving an independent variable for the specific enthalpy of the fluid to tackle challenging non-isothermal dynamics.
Since the focus is on coupling with thermodynamics, for simplicity we consider only the pressure field as the force driving the flow.
For inclusion of gravity, Fickian diffusion, or mechanical dispersion, see \citet{moortgat2011} for example.
\subsection{Thermal compositional multiphase flow}\label{subsection:flow-transport}
Consider a fluid mixture consisting of $\eta = 1 \dots \ncomp$ components and $\gamma = 1 \dots\nphase$ phases in a porous medium.
We assume that the solid skeleton of the porous medium is fixed, restrict our considerations to fluid phases only, gas and liquid, and anticipate the number $\nphase$ as the maximum number of possible phases for a target range of pressure and temperature values.
The balance of mass for a component $\xi$ in phase $\delta$ is given by
\begin{equation}\label{equ:mass-balance-comp-in-phase}
    \pdv{}{t}\phi\rho_{\delta}s_{\delta}x_{\xi\delta}
    + \nabla\cdot\left(x_{\xi\delta}\rho_{\delta}\mathbf{v_{\delta}}\right)
    =
    \sum_{\gamma\neq\delta}q_{\xi\gamma} + r_{\xi\delta},
\end{equation}
with $\phi$ denoting porosity, $\rho_\delta$ and $s_\delta$ the density and saturation of phase $\delta$ respectively, $x_{\xi\delta}$ the partial fraction of mass of component $\xi$ in phase $\delta$, $q_{\xi\gamma}$ a source term due to mass exchange between phases and $r_{\xi\delta}$ a local source of mass.
The volumetric flux $\mathbf{v_{\delta}}$ of phase $\delta$ is modeled using Darcy's law
\begin{equation}\label{equ:darcy-flux}
    \mathbf{v_\delta} = -\dfrac{k_{\delta}}{\mu_{\delta}}\absperm\nabla p,
\end{equation}
where we neglect gravity, diffusion of mass, and capillarity;
$k_\delta$ denotes the relative permeability and $\absperm$ the absolute permeability;
$\mu_\delta$ denotes the dynamic viscosity of phase $\delta$ and $p$ the pressure.

The \textit{component mass balance} for a component $\xi$ is obtained by summing \Cref{equ:mass-balance-comp-in-phase} over the context of phases:
\begin{equation}\label{equ:mass-balance-comp}
    \pdv{}{t}\phi\rho z_{\xi}
    +\nabla\cdot\left(\sum_{\gamma}x_{\xi\gamma}\rho_{\gamma}\mathbf{v_\gamma}\right)
    =
    r_{\xi}.
\end{equation}

We note a couple of steps required to arrive at \Cref{equ:mass-balance-comp}.
First, we exploited the definitions 
\begin{subequations}\label{equs:defs-mass-balance-comp}
    \begin{align}
        y_{\delta} &= \dfrac{\rho_{\delta}}{\sum_{\gamma}\rho_{\gamma}s_{\gamma}} s_{\delta}, \label{equ:def-phase-fraction}\\
        z_{\xi} &= \sum_{\gamma}y_{\gamma}x_{\xi\gamma},\label{equ:def-overall-fraction}\\
        \rho &= \sum_{\gamma}\rho_{\gamma}s_{\gamma},\label{equ:def-mixture-density}
    \end{align}
\end{subequations}
with $y_\delta$ denoting the fraction of overall mass in phase $\delta$, $z_\xi$ the fraction of overall mass of component $\xi$, and $\rho$ the density of the fluid mixture as a sum of phase densities weighed with the fraction of pore volume they occupy.

Second, the source terms due to mass exchange between phases cancel due to reciprocity:
\begin{equation}
    \sum_{\delta} \sum_{\gamma\neq\delta}q_{\xi\gamma} = 0.
\end{equation}
And third, fractional quantities fulfill the unity constraint.
For $\varphi_\circ \in \{z_\circ, y_\circ, s_\circ, x_{\circ 1},\dots, x_{\circ \nphase}\}$ we have
\begin{equation}\label{equ:unity-of-fractions}
    1 = \sum_j \varphi_j,
\end{equation}
with the index $j$ being in the context of components, phases or components in a phase respectively.
In particular, for $\ncomp$ fluid components we require \Cref{equ:mass-balance-comp} and variables $z$ only for $\xi = 2 \dots \ncomp$, where we set the first component as the reference component without loss of generality.
Analogously, we express $s_1$ and $y_1$ by unity of fractions.

Whichever context we first eliminate, phases or components, the option to eliminate both is viable and required.
By doing so, we obtain the balance of total fluid mass, or \textit{pressure equation}
\begin{equation}\label{equ:pressure-equation}
    \pdv{}{t}\phi\rho
    +\nabla\cdot\left(\sum_{\gamma}\rho_{\gamma}\mathbf{v_{\delta}}\right)
    =
    r,
\end{equation}
which is the governing equation for our pressure variable $p$.
The term $r=\sum_\xi r_\xi$ represents the total mass or pressure source term.

As the last governing equation for flow and transport, we introduce the balance of total energy
\begin{align}
    &\pdv{}{t}\left(\phi(\specEnthalpy\rho - p) + (1-\phi)u_s\rho_s\right) \nonumber\\
   & +\nabla\cdot\left(\sum_{\gamma} \specEnthalpy_\gamma\rho_{\gamma}\mathbf{v_{\delta}}
    - \left(\phi \left(\sum_\gamma s_\gamma \kappa_\gamma\right) + (1-\phi)\kappa_s\right)\mathbf{I}\nabla T
    \right)  \nonumber\\
    &=
    r_e, \label{equ:total-energy-balance}
\end{align}
where $u_s,\rho_s,\kappa_s$ represent the internal energy, density and thermal conductivity of the porous medium, and $\specEnthalpy_\gamma$, $\kappa_\gamma$ the specific enthalpy and thermal conductivity of a fluid phase.
The diffusive flux of energy is modeled by Fourier's law.
The energy source term $r_e$ specifies the energy entering the system through the mass source $r$.
Other heat sources can be added if necessary.

\Cref{equ:total-energy-balance} is obtained analogously to \Cref{equ:pressure-equation} by introducing a basic balance equation of energy for a fluid component in a phase, as well as the balance of energy in the porous medium, summing over the component and phase context and assuming local thermal equilibrium; i.e., assuming that all phases in both fluid and porous medium have an equal temperature $T$.
However, note that we explicitly introduce the specific enthalpy of the fluid mixture $\specEnthalpy$ as a variable, while keeping $T$.
This modeling choice will be motivated by the local equilibrium formulation in \Cref{subsection:phase-equilibrium}.

\Cref{table:balance-equations} summarizes the governing equations for flow and transport, assuming a multiphase, multicomponent fluid, and a single solid phase in the porous medium. Note that due to the dependency between the two energy variables $\specEnthalpy$ and $T$, the energy equation is both advective and diffusive. The pressure equation, on the other hand, is purely diffusive in $p$ and the component mass balance equations are purely advective in $z_\xi$.

We conclude the modeling of flow and transport by defining the vector of primary variables, or transport variables, as
\begin{equation}\label{equ:def-primary-variables}
    \varX = [p, h, z_2,\dots,z_{\ncomp}].
\end{equation}

\begin{table}
\centering
\begin{tabular}{@{}l l l@{}}
\toprule
\textbf{Balance equation} & \textbf{Definition} & \textbf{Type} \\
\midrule
pressure equation
&
$\pdv{}{t}\phi\rho-\nabla\cdot\left(\lambda\absperm\nabla p\right)=r$
& diffusive \\
\addlinespace
\makecell[l]{$\ncomp -1$ component\\mass balance}
&
$\pdv{}{t}\phi\rho z_\xi -\nabla\cdot\left(\lambda_\xi \absperm\nabla p\right)=r_\xi$
& advective \\
\addlinespace
energy balance
&
$\pdv{}{t}\left(\phi(\specEnthalpy\rho - p) + (1-\phi)u_s\rho_s\right) - \nabla \cdot \left(\lambda_e \absperm\nabla p + \mathbf{q_e}\right) = r_e$
& \makecell[l]{advective-\\diffusive} \\
\midrule
\textbf{Constitutive relations} & \textbf{Definition} & \\
\midrule
component mass mobility & $\lambda_\xi = \sum_{\gamma} x_{\xi\gamma}\rho_{\gamma} \dfrac{k_\gamma}{\mu_\gamma}$ & \\
\addlinespace
total mass mobility & $\lambda = \sum_\xi \lambda_\xi = \sum_{\gamma}\rho_{\gamma} \dfrac{k_\gamma}{\mu_\gamma}$ & \\
\addlinespace
energy mobility & $\lambda_e = \sum_{\gamma} \specEnthalpy_\gamma\rho_{\gamma} \dfrac{k_\gamma}{\mu_\gamma}$ & \\
Fourier's law & $\mathbf{q_e} = \left(\phi \left(\sum_\gamma s_\gamma \kappa_\gamma\right) + (1-\phi)\kappa_s\right)\mathbf{I}\nabla T$ & \\
\addlinespace
\bottomrule
\end{tabular}
\caption{\label{table:balance-equations} Governing equations for compositional multiphase flow and the nature of the transport it models. The associated primary variables are $p$, $z_\xi$ and $\specEnthalpy$ respectively.}
\end{table}

\subsection{Local equilibrium model}\label{subsection:phase-equilibrium}
\Cref{subsection:flow-transport} introduced a single pressure and temperature variable, touching on the notion of local equilibrium.
It also introduced phase fractions, saturations and partial fractions, as well as two energy-related variables, leaving the mathematical models summarized in \Cref{table:balance-equations} open.
This section presents the local phase equilibrium (sub-) problem and effectively closes the coupled-physics model. We chose the equilibrium formulation in terms of enthalpy because of its suitability for narrow-boiling fluids \cite{zhu2014}.

We first define the vector of secondary variables, or local variables, that complements the primary transport variables given in \Cref{equ:def-primary-variables}.
They are given by
\begin{equation}\label{equ:def-secondary-variables}
    \varY = [T, s_2,\dots,s_{\nphase}, y_2, \dots, y_{\nphase}, x_{11}, \dots, x_{\ncomp\nphase}].
\end{equation}
Note that the first saturation and phase fraction variables are eliminated as per \Cref{equ:unity-of-fractions}.
Saturations are usually not required in the local problem, since they can be obtained using \Cref{equ:def-phase-fraction}, but we include them for completeness.

The local problem addresses the issue of phase appearance and disappearance.
The thermodynamically consistent way of approaching this problem is to define the target state of the fluid, choose the correct state function, and minimize it \cite{michelsen2004}.
For given values of $p, \specEnthalpy$ and $z_\xi$, the respective minimization problem is given by \cite{michelsen1999}
\begin{equation}\label{equ:ph-minimization-problem}
    \varY^\star = \underset{\varY}{\arg\min} ~\frac{1}{T}\left(\specGibbs(\varY) - h\right),
\end{equation}
where $\specGibbs$ denotes the specific Gibbs energy of the fluid.
The minimization problem \eqref{equ:ph-minimization-problem} is subject to local mass constraints given by \Cref{equ:def-overall-fraction}.

The number of phases $\nphase$ actually present at the minimum is a priori unknown.
Instead of performing phase stability calculations to determine $\nphase$ \cite{michelsen1982-1,nichita2024}, we opt for the persistent-variable formulation of the problem \cite{gupta1990,gharbia2021}.
This approach fixes the maximum possible number of phases and binds the phase fractions from below by zero:
\begin{equation}\label{equ:y-nonnegative}
    y_\delta \geq 0,~\forall \delta\in\{1,\dots,\nphase\}.
\end{equation}

\citet{gharbia2021} show that for isothermal equilibrium definitions in terms of $p$ and $T$, where the objective function is the specific Gibbs energy $\specGibbs$, the minimization problem with constraints \eqref{equ:def-overall-fraction} and \eqref{equ:y-nonnegative} is well-defined and has an unique solution which corresponds to the physical solution.
This requires the introduction of \textit{extended partial fractions} $\extfrac_{\xi,\delta}$ for cases where $y_\delta =0$, as well as the equality of chemical potentials for vanished phases.
The relation between extended partial fractions and (physical) partial fractions is as follows.
\begin{equation}\label{equ:partial-fraction-relation}
    x_{\xi\delta} = \begin{cases}
    \extfrac_{\xi\delta} & ,~y_\delta > 0, \\
    \dfrac{\extfrac_{\xi\delta}}{\sum_\eta \extfrac_{\eta\delta}} & ,~y_\delta = 0.
    \end{cases}
\end{equation}

An extension of this approach to include other equilibrium definitions, including isenthalpic ones in terms of $p$ and $\specEnthalpy$, is possible \cite{lipovac2024}.
Expressing the Gibbs energy in terms of chemical potentials, and equivalently as fugacities, and using Lagrangian techniques to obtain first-order conditions for a minimum, we arrive at a system of equations of the form
\begin{equation} \label{equ:ph-system}
    \begin{array}{r l}
        \begin{bmatrix}
            \Lambda(\varY) \\
            \Upsilon(\varY) \\
            \Gamma(\varY)\odot\LagMultipier(\varY)
        \end{bmatrix} &= 0~,  \\[4ex]
        \Gamma(\varY)~,~\LagMultipier(\varY) & \geq 0~,
    \end{array}
\end{equation}
where
\begin{equation}\label{equ:def-equilibrium-system-terms}
    \Lambda(\varY) =
    \begin{bmatrix}
        \vdots \\
         \big(\extfrac_{\xi\delta}~\fugCoeff_{\xi\delta}
        - \extfrac_{\xi1}~\fugCoeff_{\xi1}\big)_{\xi\geq 1, \delta \geq 2} \\
        \vdots \\
        \big(z_\xi - \sum\limits_\gamma y_\gamma \extfrac_{\xi\gamma}\big)_{\xi\ge 2} \\ 
        \vdots
    \end{bmatrix},~~
    \Upsilon(\varY) = 
    \begin{bmatrix}
        h - \sum_\gamma y_\gamma h_\gamma
    \end{bmatrix},~~
    \Gamma(\varY) =
    \begin{bmatrix}
        1 - \sum_{\gamma \geq2} y_\gamma \\ y_2\\\vdots \\ y_{\nphase}
    \end{bmatrix},~~
    \LagMultipier(\varY) = 
    \begin{bmatrix}
        1 - \sum\limits_\eta \extfrac_{\eta1} \\
        \vdots \\
        1 - \sum\limits_\eta \extfrac_{\eta\nphase}
    \end{bmatrix}.
\end{equation}
The component-wise product is denoted by $\odot$, and $\fugCoeff_{\xi\delta}$ is the fugacity coefficient of component $\xi$ in phase $\delta$.

In particular, the equations for the persistent-variable $pT$-equilibrium \cite{gharbia2021} are a subset of the System \eqref{equ:ph-system}, with $\Upsilon$ being the addition which constrains the mixture enthalpy $\sum_\gamma y_\gamma \specEnthalpy_\gamma$  to a given $\specEnthalpy$ and allows $T$ to be an additional degree of freedom.
This allows for an efficient implementation where we either solve the $pT$ problem and update $\specEnthalpy$, or solve the $p\specEnthalpy$ problem as a whole where $pT$ is expected to fail \cite{zhu2016}.
Furthermore, \Cref{equ:ph-system} closes the transport system introduced in \Cref{subsection:flow-transport} since the enthalpy constraint $\Upsilon$ relates the two energy variables $\specEnthalpy$ and $T$ to each other.
Through this approach and by explicitly defining a variable $\specEnthalpy$, we now have a model capable of tackling challenging boiling and condensation scenarios in the compositional flow simulation.
\subsection{Model summary}\label{subsection:model-summary}
The system derived in \Cref{subsection:flow-transport,subsection:phase-equilibrium} represents a coupled physics system with transport of mass, energy and thermodynamics of phase change.
For the $2 + \ncomp -1$ flow and transport variables, $p, \specEnthalpy$ and $z_\xi,~\xi\in\{2,\dots,\ncomp\}$, we introduce the $2 + \ncomp -1$ PDEs
\begin{subequations}\label{equ:pde-subsystem}
    \begin{align}
        \pdv{}{t}\phi\rho-\nabla\cdot\left(\lambda \nabla p\right) - r &=0,
        \label{equ:pde-subsystem-p}\\
        \pdv{}{t}\phi\rho z_\xi -\nabla\cdot\left(\lambda_\xi \nabla p\right) - r_\xi &=0,~\xi\in\{2,\dots,\ncomp\}
        \label{equ:pde-subsystem-z}\\
        \pdv{}{t}\left(\phi(h\rho - p) + (1-\phi)u_s\rho_s\right) - \nabla \cdot \left(\lambda_e \nabla p + \mathbf{q_e}\right) - r_e &=0.
        \label{equ:pde-subsystem-h}
    \end{align}
\end{subequations}

For the $\nphase - 1 + \ncomp\nphase$ phase fractions and extended partial fractions, $y_\delta,~\delta\in\{2,\dots,\nphase\}$, and $\extfrac_{\xi,\delta},~\xi\in\{1,\dots,\ncomp\},~\delta\in\{1,\dots,\nphase\}$, and temperature $T$, we introduce the $p\specEnthalpy$-equilibrium system, a set of algebraic equations (AE) of form
\begin{subequations}\label{equ:equilibrium-subsystem}
    \begin{align}
        \Lambda &=0,
        \label{equ:equilibrium-subsystem-isofug} \\
        \Upsilon &=0,
        \label{equ:equilibrium-subsystem-h} \\
        \min\left\{\Gamma,\LagMultipier\right\} &= 0,
        \label{equ:equilibrium-subsystem-complementarity}
    \end{align}
\end{subequations}
where the complementarity conditions are written in semi-smooth form.
With $\Lambda$ containing $\ncomp(\nphase - 1) +\ncomp - 1$ equations, $\Upsilon$ one, and the $\nphase$ complementarity conditions, this subsystem is also closed.
For the relation between saturations and phase fractions, \Cref{equ:def-phase-fraction} is used, and for the relation between extended and physical partial fractions, \Cref{equ:partial-fraction-relation} is used.
They can either be appended directly for global closure, or used to eliminate the phase fractions and physical partial fractions.
In any case, with the exception of the $2 + \ncomp -1$ flow and transport variables, we understand all other degrees of freedom as \textit{local variables}.
The locality is understood considering the above distinction of PDE and AE subsystem.
Furthermore, for every set of flow and transport variables $(p, h, z_\xi)$ there exists a unique solution of the local equilibrium problem, which are the phase separation variables.
We aim to exploit this relation in the nonlinear solution method.

\section{Numerical methodology}\label{section:numerical-method}
The system of equations at hand represents an advection-diffusion problem with local non-smooth dynamics in the form of phase equilibrium calculations.
Realistic constitutive modeling using an equation of state for the fluid additionally introduces challenging nonlinearities.
While there is extensive literature on how to deal with System \eqref{equ:pde-subsystem} and its individual equations, we focus in this paper on the inclusion and exploitation of numerical techniques for the equilibrium Subsystem \eqref{equ:equilibrium-subsystem}.

In \Cref{subsection:discretization} we briefly summarize the discretizations used and introduce the system in discrete form.
In \Cref{subsection:nonlinear-solver} we present a nonlinear solution strategy which includes a nested algorithm for the equilibrium problem.
The base method is a global semi-smooth Newton algorithm with line search.
We introduce an equilibration step in between iterations, which provides an intermediate update for the secondary variables $\varY$.
The new strategy results in an efficient and robust nonlinear solver, which enables us to conduct simulations with time steps on the scales we are interested in.
\subsection{Discretization}\label{subsection:discretization}
\Cref{equ:pde-subsystem,equ:equilibrium-subsystem} are a system of coupled partial differential and algebraic equations.
For spatial discretization of \Crefrange{equ:pde-subsystem-p}{equ:pde-subsystem-h} we use a finite-volume approach to discretization with scalar values per cell centers for all variables in $\varX$ and $\varY$.
Fluxes on cell faces are approximated using a multi-point flux approximation \cite{aavatsmark2002}, for both Darcy and Fourier flux.
Nonlinear weights in the advective fluxes on faces are approximated using first-order upwinding \cite{courant1952}.
For the temporal discretization we use the backward Euler method in a fully implicit scheme.
The algebraic \Crefrange{equ:equilibrium-subsystem-isofug}{equ:equilibrium-subsystem-complementarity} are solved as given in cell centers only.
A forward automatic differentiation framework is used to evaluate the derivatives.
The derivatives of thermodynamic properties of phases are evaluated using the underlying equation of state.
The implementation is done in \textit{PorePy} \cite{keilegavlen2020}.

All steps considered, at a given time $t_n\in[0,t_{end}]$ and $\varX_n = \varX(t_n), \varY_n = \varY(t_n),$, and for given $\varX_{n-1}, \varY_{n-1}$, the system of nonlinear equations to be solved can be written as
\begin{equation}\label{equ:discrete-system}
    \sysF(\varX_n,\varY_n) = \begin{bmatrix}
        \resft(\varX_n,\varY_n) \\
        \resph(\varX_n,\varY_n)
    \end{bmatrix}
    = 0,
\end{equation}
with $\resft$ denoting the discretized flow and transport equations and $\resph$ the local equilibrium equations. The non-smoothness is contained within $\resph$.

\begin{remark}
    When discretizing balance equations, volume discrepancies can occur between the actual fluid volume in the pore space, which depends on the grid discretization and the measure of performed volume integrals, and the volume as calculated by an equation of state, which is usually in SI units.
    This discrepancy requires a scaling of terms that contain densities, effectively correcting the discrepancy using pressure \cite{weiss2014,trangenstein1989}.
    The correction is usually performed on the continuous level as a general solution independent of the grid or used discretizations.
    For the sake of the methodology presented in this paper, we circumvent this issue by formulating and discretizing the entire problem in base SI units, most importantly in meters and cubic meters.
    We also use finite-volume methods, where the measure of the volume integral is in meters.
    Therefore, there are no discrepancies in volume as assumed in the PDEs and in the equilibrium problem.
\end{remark}
\subsection{Nonlinear solution strategy}\label{subsection:nonlinear-solver}
Following a time-stepping scheme, \Cref{equ:discrete-system} can in principle be solved using a semi-smooth Newton method:
\begin{equation}\label{equ:newton-system}
    \begin{bmatrix}
        \jacof{\varX} \resft(\varX_{n}^{i-1},\varY_{n}^{i-1}) & \jacof{\varY} \resft(\varX_{n}^{i-1},\varY_{n}^{i-1}) \\
        \jacof{\varX} \resph(\varX_{n}^{i-1},\varY_{n}^{i-1}) & \jacof{\varY} \resph(\varX_{n}^{i-1},\varY_{n}^{i-1})
    \end{bmatrix}
    \begin{bmatrix}
        \Delta \varX_{n}^{i} \\ \Delta\varY_{n}^{i}
    \end{bmatrix}
     = - \begin{bmatrix}
         \resft(\varX_{n}^{i-1},\varY_{n}^{i-1}) \\
         \resph(\varX_{n}^{i-1},\varY_{n}^{i-1})
     \end{bmatrix},
\end{equation}
with superscript $i$ denoting the iteration and $\jacof{x} F$ the partial Jacobian of $F$ with respect to $x$.
\citet{lauser2011} present results using a $pT$ formulation and heuristic laws for thermodynamic properties.
However, basing the thermodynamic model of the fluid on an equation of state introduces more challenging nonlinearities and phase behavior.
In addition, high-enthalpy scenarios and rapid injection and production cycles burden the nonlinear solver.
In order to converge, the time step size $\Delta t$ must be drastically decreased, rendering simulations over a meaningful time span unfeasible.
Therefore, a solution strategy is required that directly addresses the local problem.

We want to exploit the formulation of the equilibrium problem, which is now formulated as a closed subsystem $\resph$.
For every $\varX$ (transported mass and energy; i.e., thermodynamic target state), there exists a unique $\varY$ that solves the local equilibrium such that $\resph(\varX,\varY) =0$.
The main idea is to solve the thermodynamic subproblem in the course of the iterations up to a desired precision, and do so efficiently by exploiting its local nature.
After performing a Newton iteration $i$ and obtaining a global update, we apply another solver to obtain an improved update for $\varY$:
\begin{align}
    \varX_n^i &= \varX_n^{i - 1} + \Delta \varX_n^i,~~\varY_n^i = \varY_n^{i - 1} + \Delta \varY_n^i \nonumber \\
    \varY_n^i &\leftarrow \hat{\varY}_n^i = \hat{\varY}_n^i(\varX_n^i), ~~\text{such that} \label{equ:local-update-Y}\\
    \lVert \resph(\varX_n^i, \hat{\varY}_n^i)\rVert & \leq \epsilon_1. \nonumber
\end{align}
Notably, by doing so we ensure that the local equilibrium assumption is fulfilled also in the course of iterations.

In this work, we rely on a non-parametric interior point method \cite{vu2021} as the local solver.
Note, however, that it can be any optimization algorithm that solves Problem \eqref{equ:ph-minimization-problem}.
$\resph$ represents the first-order conditions of the said optimization problem and any solution to it will lead to $\resph$ going to zero.
Moreover, due to the locality of the thermodynamic problem, this can be done in parallel over the cells in the computational domain.
To further increase the efficiency of this approach and avoid solving the local problem in cells where mass and energy do not change significantly ($\Delta\varX_n^i\approx 0$) , we can use $\varY_n^i$ as the initial guess, which will lead the local solver to exit quickly.

We can interpret this approach in two ways.
One is a type of splitting algorithm that resolves the disparate time scales of coupled physics.
While flow and transport proceed in some characteristic time, the phase equilibrium problem is instantaneous.
Keeping physics on an instantaneous time scale resolved at all times and iterations supports the convergence of the global Newton solver.
The other is to view the application of the local solver to obtain $\hat{\varY}_n^i$ as a sophisticated correction of the Newton update.

In either way, the global Newton iteration frequently violates not only the local equilibrium assumption, but also the boundedness of fractional variables to the interval $[0,1]$.
To counteract those violations, a line search algorithm is required to scale Newton updates with some factor $\alpha$, such as the Armijo line search \cite{armijo1966}.
Furthermore, a post-processing of the values $\varX_n^i,\varY_n^i$ is required, projecting the values of fractional variables back to $[0,1]$ and re-normalizing them such that the unity constraint \eqref{equ:unity-of-fractions} is fulfilled.
This is done locally.

\begin{figure}
    \centering
    \begin{tikzpicture}[node distance=1.2cm]

    \node (setupnode) [process] {\textbf{setup}};
    \node (initnode) [process, below of=setupnode] {initialize $p, T, z_\xi$};
    \node (initflashnode) [process, below of =initnode] {initial equilibration};

    \node (bcnode) [process, right of=initflashnode, xshift=1cm] {update BC};

    \node (linsolvenode) [process, right of=bcnode, xshift=1cm] {lin. solve};
    \node (postprocessnode) [process, above of=linsolvenode, yshift=1.6cm] {postprocess};
    \node (equilibratenode) [decision, right of=postprocessnode, xshift=2cm] {equilibrate};
    \node (flashnode) [process, below of=equilibratenode, xshift=-1cm, yshift=-0.2cm, fill=black!2] {\textbf{local equilibration}};
    \node (noflashnode) [process, below of=equilibratenode, xshift=1cm, yshift=-0.2cm] {update fluid properties};
    \node (rediscfluxnode) [optprocess, below of=equilibratenode, yshift=-1.6cm] {re-discretize fluxes};
    \node (rediscupwindnode) [process, right of=rediscfluxnode, xshift=2cm] {re-discretize upwinding};
    \node (convergednode) [decision, above of=rediscupwindnode, yshift=0.2cm] {converged};
    \node (maxiternode) [decision, above of=convergednode, yshift=0.2cm] {max. iter.};
    \node (updatetimenode) [process, right of=maxiternode, xshift=1.5cm] {update $t, \Delta t$};
    \node (endtimenode) [decision, right of=convergednode, xshift=1.5cm] {max. time};
    \node (stopnode) [process, right of=rediscupwindnode, xshift=1.5cm] {\textbf{stop}};

    \draw [arrow] (setupnode) -- (initnode);
    \draw [arrow] (initnode) -- (initflashnode);
    \draw [arrow] (initflashnode) -- (bcnode);
    \draw [arrow] (bcnode) -- (linsolvenode);
    \draw [arrow] (linsolvenode) -- (postprocessnode);
    \draw [arrow] (postprocessnode) -- (equilibratenode);
    
    \draw [arrow] (equilibratenode) --node[anchor=east] {yes}  (flashnode);
    \draw [arrow] (equilibratenode) --node[anchor=west] {no}  (noflashnode);
    \draw [arrow] (flashnode) -- (rediscfluxnode);
    \draw [arrow] (noflashnode) -- (rediscfluxnode);
    \draw [arrow] (rediscfluxnode) -- (rediscupwindnode);
    \draw [arrow] (rediscupwindnode) -- (convergednode);

    \draw [arrow] (convergednode) --node[anchor=south] {yes}  (endtimenode);
    \draw [arrow] (convergednode) --node[anchor=west] {no}  (maxiternode);
    \draw [arrow] (endtimenode) --node[anchor=west] {no}  (updatetimenode);
    \draw [arrow] (endtimenode) --node[anchor=west] {yes}  (stopnode);

    \draw [arrow] (maxiternode) --node[anchor=south] {yes}  (updatetimenode);
    \draw[thick] (maxiternode) --node[anchor=south west]{no} (10,1);
    \draw[thick] (10,1) -- (3.4,1);
    \draw[thick] (3.4,1) -- (3.4,-1.8);
    \draw[arrow] (3.4,-1.8) -- (linsolvenode);

    \draw[thick] (updatetimenode) -- (12.5,1.3);
    \draw[thick] (12.5,1.3) -- (2.2,1.3);
    \draw[arrow] (2.2,1.3) -- (bcnode);

    \node[rectangle, draw, minimum width=2cm, minimum height=3.5cm] (rectinit) at (0,-1.2) {};
    \node[thick, yshift=-0.3cm] at (rectinit.south) {\small \textbf{Initialization}};

    \node[rectangle, draw, minimum width=8.6cm, minimum height=4.1cm] (rectnewton) at (7.6,-0.9) {};
    \node[thick, yshift=-0.3cm] at (rectnewton.south) {\small \textbf{Newton step}};

    \end{tikzpicture}
    \caption{Flow chart for nonlinear solver. The Newton algorithm applied to each time step forms the basis. Within each iteration, the solution strategy is adapted to use the local equilibrium solver and feed the results back into the global system.}
    \label{figure:solver-flow-chart}
\end{figure}
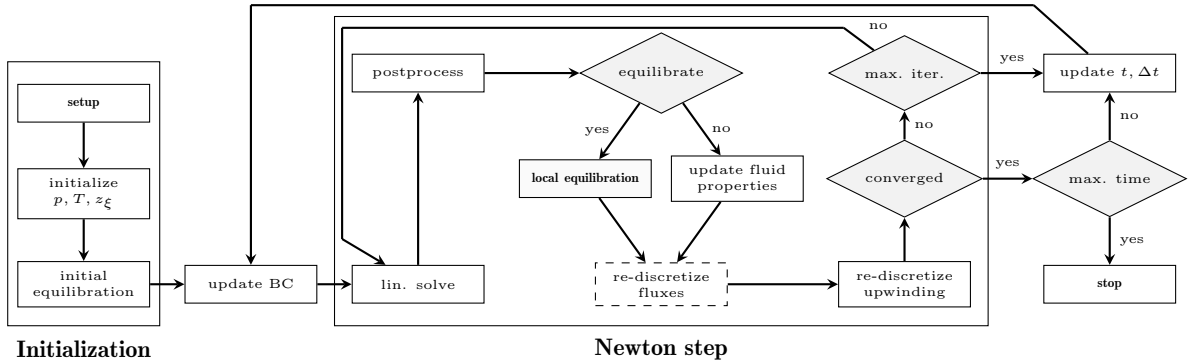

\Cref{figure:solver-flow-chart} visualizes the proposed nonlinear solution strategy.
We provide more details about the procedure below.

\begin{enumerate}
\setlength{\itemsep}{-5pt}
    \item \textbf{setup}:
    Basic simulation setup, creating the grid, variables equations etc.
    \item\textbf{initialization}:
    Despite the large amount of variables, it suffices to define the initial state of the fluid in terms of a target equilibrium state.
    We chose an initialization in terms of pressure, temperature and mass since they are a natural choice for engineering.
    All other variables, including the specific enthalpy $\specEnthalpy$ are initialized using the equilibration results.
    \item\textbf{update boundary conditions}:
    The underlying equation of state for the fluid is exploited to provide consistent boundary conditions without significant effort.
    When Dirichlet-type conditions for pressure and temperature are provided, equilibration on the boundary will yield the corresponding values for the mass mobilities on the boundary.
    When providing Neumann-type data, the only thing to consider is that the total mass in- and outflow on the boundary must be equal to the sum of component mass in- and outflow.
    \item\textbf{linear solver}:
    We apply a Schur complement technique using the PDEs and variables $\varX$ as the primary block.
    The local equations and variables represented by $\jacof{\varY} \resph$ make a block-diagonal matrix, with blocks of size $\nphase(\ncomp + 1)$ per cell.
    It is therefore efficiently invertible and we construct the reduced system
    \begin{align}
        \left(\jacof{\varX} \resft - \jacof{\varY} \resft(\jacof{\varY} \resph)^{-1}\jacof{\varX} \resph\right) \cdot\Delta\varX_n^i &=
        - \resft + \jacof{\varY} \resft(\jacof{\varY} \resph)^{-1}\resph
        , \label{equ:reduced-linear-system} \\ 
        \Delta\varY_n^i&= (\jacof{\varY} \resph)^{-1}\left(\resph - \jacof{\varX} \resph \Delta \varX_n^i\right). \nonumber
    \end{align}
    We use a direct solver to solve \Cref{equ:reduced-linear-system}.
    With an increasing refinement, iterative solvers are required for efficiency reasons.
    An approach to iterative solver for coupled problems in porous media flow is given by \citet{roy2020}.
    \item\textbf{postprocessing}:
    Safeguard the unity of fractions.
    The equilibrium problem is not defined for $z_\xi\notin [0,1]$, and using other fractions with values outside the interval as an initial guess for the local minimization problem is likely to result in failure.
    The postprocessing step can also include a line search algorithm or Newton chopping update, such as the Appleyard chop for the saturation updates \cite{younis2009}.
    \item\textbf{local equilibration}:
    This decision leaves room for tailoring the solution strategy to individual simulations.
    For example, we can perform the equilibration only every couple of iterations.
    We can also chose which equilibration to perform:
    A $pT$-equilibrium with an update to $\specEnthalpy$ locally, or a $p\specEnthalpy$-equilibrium including an update for $T$.
    In any case, the equilibration updates the fluid properties as well as all other variables not included in the target equilibrium state; i.e., $\hat{\varY}_n^i$.
    If no equilibration is requested, a simple update of fluid properties using currant iterate values of $\varX_n^i, \varY_n^i$ is performed.
    \item\textbf{re-discretize fluxes}:
    This step is required for the heat flux $\mathbf{q_e}$ as saturations and thermal conductivities change after equilibration.
    It is also required if the total mass mobility $\lambda$ is taken as a nonlinear contribution to the diffusive tensor in \Cref{equ:pde-subsystem-p} \cite{duran2025}.
    \item\textbf{re-discretize upwinding}:
    This step is required for any transport problem in case the direction of the flux changes.
    \item\textbf{convergence check}:
    We perform a residual-based convergence check; i.e., we check if the Euclidean norm of the residual of each equation individually falls below the target threshold $\epsilon$.
    With the exception of the isofugacity constraints in \Cref{equ:equilibrium-subsystem-isofug}, all equations are a form of mass and energy conservation, either locally or non-locally.
    This criterion fits the requirements of a conservative algorithm.
    The isofugacity constraints require a special treatment.
    As highlighted by \citet{iapws1997}, it is a challenge to devise thermodynamic models where the fugacities of components in the multiphase region are equal for a given bit precision.
    This issue is aggravated numerically due to the exponential function being part of the expression for $\fugCoeff_{\xi\delta}$.
    Therefore we use a relaxed condition for these local equations, requiring their residual to fall below $\num{1e-2}$ only.
    This relaxation allows an error of up to 1\% in terms of how the mass of a component is distributed across different phases, but it does not violate the conservation of total mass of components.
    \item\textbf{maximum iteration check}:
    If the maximum number of global iterations is reached, the model is flagged as non-convergent and a reduction of the time step size is triggered.
    \item\textbf{update $t,\Delta t$}:
    We use a simple adaptive time stepping scheme, with a relaxing and a constraining factor $\Delta t_n = c \Delta t_{n -1}$ depending on the number of iterations required to converge.
    If the model is flagged as non-convergent or diverges, the time step size is reduced.
    If a defined number of time step size reductions at a time $t$ is reached, the model exits the simulation unsuccessfully.
    \item\textbf{maximum time $t$ reached}:
    Condition for stopping the simulation and exiting successfully.
\end{enumerate}

\begin{remark}
    The assembly of the Jacobian in \Cref{equ:newton-system} leaves space for various techniques.
    We use a combination of an internal automatic differentiation framework \cite{keilegavlen2020} to assemble the whole system, and symbolic computing of thermodynamic properties and their derivatives, with a subsequent compilation for efficient and parallel evaluation.
    In this way, the assembled and inverted Jacobian matrices \eqref{equ:newton-system}-\eqref{equ:reduced-linear-system} are exact.
\end{remark}

\section{Results}\label{section:results}
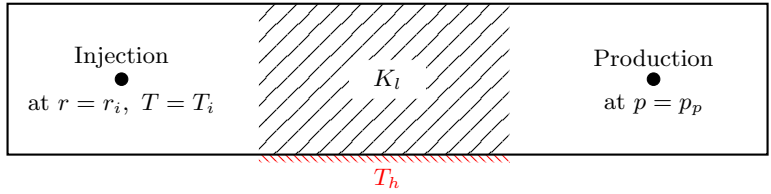
\begin{figure}[H]
    \centering
    \begin{tikzpicture}
    \draw[color=white, pattern={Lines[angle=45,distance=0.2cm]}, pattern color=black] (3.3,0) rectangle (6.6,2);
    \draw[color=white, pattern=north west lines, pattern color=red,] (3.3,-0.1) rectangle (6.6,0);
    \node[rectangle, thick, draw, minimum width=10cm, minimum height=2cm] (domain) at (5,1) {};
    \node[circle, draw, minimum size=1pt,scale=0.5, fill=black] (injection) at (1.5,1) {};
    \node[circle, draw, minimum size=1pt,scale=0.5, fill=black] (production) at (8.5,1) {};

    \node[yshift=0.2cm] at (injection.north) {\small Injection};
    \node[yshift=-0.25cm] at (injection.south) {\small at $r=r_i,~T=T_i$};
    \node[yshift=0.2cm] at (production.north) {\small Production};
    \node[yshift=-0.25cm] at (production.south) {\small at $p=p_p$};
    \node[yshift=-0.3cm] at (domain.south) {\small\color{red} $T_h$};

    \node[rectangle, draw, fill=white, color=white, text=black, minimum width=1cm, minimum height=0.5cm] (permbox)  at (5,1) {\small $K_l$};
    \end{tikzpicture}
    \vspace{-10pt}
    \caption{Layout of 2D simulation setup. A rectangular domain with an injection and a production well. Mass is injected at rate $r_i$ and temperature $T_i$. The pressure at production is fixed to $p_p$. The production well is modeled as a perfect sink, extracting all mass and energy flowing into respective cell. The central bottom part of the boundary has a fixed temperature $T_h$. We impose no-flow conditions everywhere else. The domain has lower permeability $K_l$ in the middle section.}
    \label{figure:simulation-setup}
\end{figure}
To test the model and methodology introduced in this work, we present a manufactured case, designed to showcase challenging physics.
We consider a 2D domain sketched in \Cref{figure:simulation-setup}, with an injector and a producer, and lower permeability in a section between the wells.
We apply Dirichlet-type boundary conditions for temperature $T_h$ at the central section at the bottom, and no-flow Neumann-type conditions on the remaining boundary.
For pressure, we apply no-flow boundary conditions on the whole boundary.
The pressure in the domain is well defined by fixing its value to $p_p$ in the production well.
In terms of the solution strategy, we present its versatility by selecting different equilibrium formulations in different parts of the domain, namely, isothermal and isenthalpic ones.
In terms of involved physics, we demonstrate that this approach is capable of handling phase transitions, narrow boiling, (de-) pressurization, nonlinear pressure evolution, and cold flooding.

As a fluid model, we use a two-phase, water-\cotwo mixture, and the Peng-Robinson equation of state as the thermodynamic model for phase properties \cite{peng1976}.
Furthermore, we assume linear relative permeabilities.
The implementation follows \citet[appendix B]{lipovac2024}, with the extension procedure introduced in \cite{gharbia2021}.
Although the Peng-Robinson EoS is not suitable for aqueous mixtures, it is sufficiently challenging in terms of nonlinearities encountered in realistic fluid modeling, and therefore serves our purposes.
An extension to other EoS for different fluid mixtures by interpolation is possible \cite{ogontula2025,driesner2007,duran2025}.
We consider two models for fluid viscosity: Constant viscosity values, equal for both phases, and the Lohrenz-Bray-Clark correlation (LBC) to obtain a viscosity dependent on
pressure, temperature and composition \cite{lohrenz1964}.

In order to first obtain the full picture, we present the simulation results using pressure and enthalpy for the local equilibration in \Cref{subsection:ph-simulation}, including an overview of the performance of the solution strategy.
In \Cref{subsection:lbc} we present a comparison of the fluid propagation and solver performance for the two viscosity models.
In \Cref{subsection:comparison} we compare the approach with a temperature-based condition and analyze the performance of the solution strategy based on the parameterization of the local solver.
The entire code is open source, and the run scripts to reproduce the simulations and figures are available online (see the declaration on data availability).
\subsection{Manufactured case under high-enthalpy conditions}
\label{subsection:ph-simulation}
\begin{figure}
    \centering
    \subfigure[]{\includegraphics[width=0.45\textwidth]{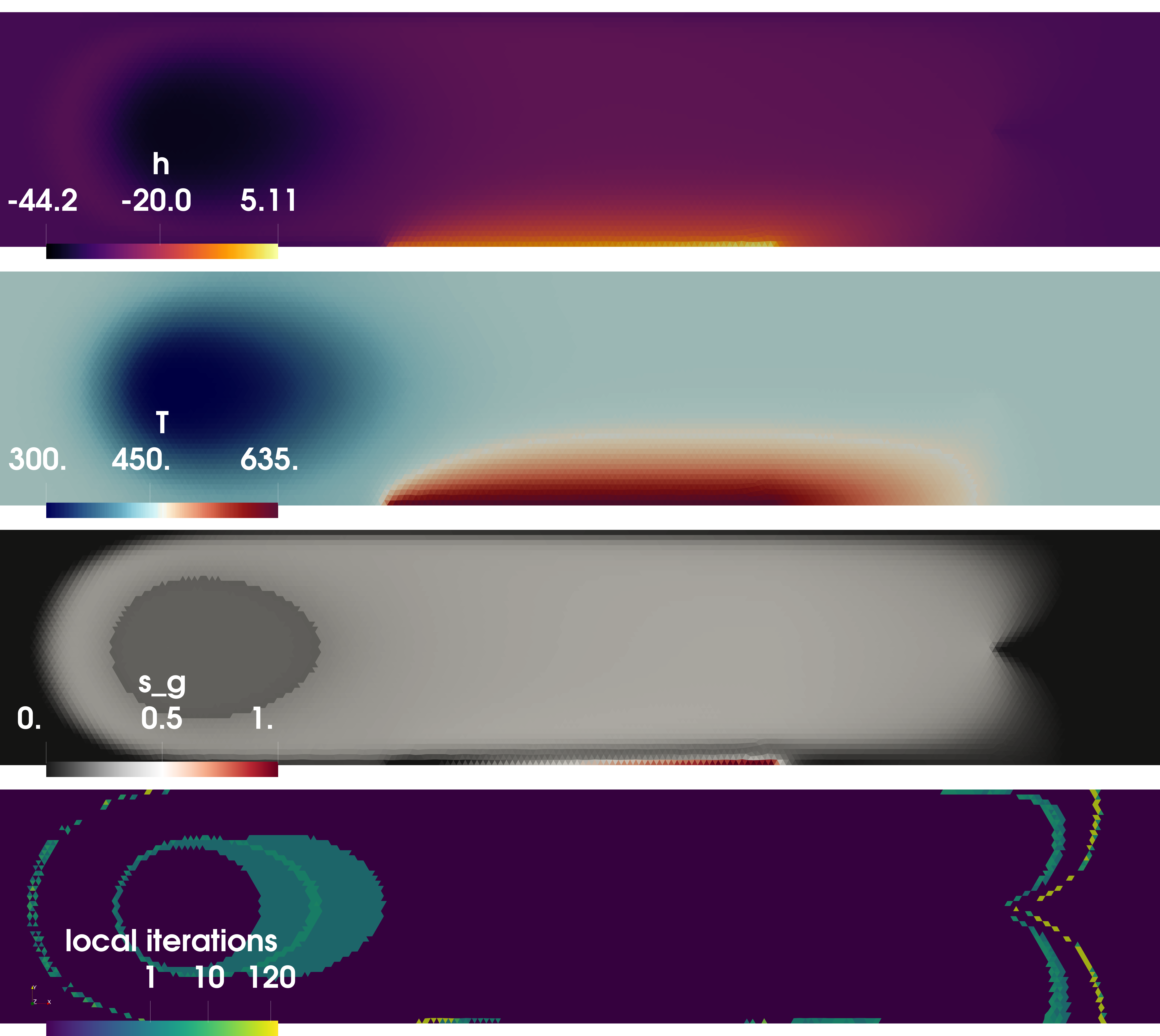}}
    \subfigure[]{\includegraphics[width=0.45\textwidth]{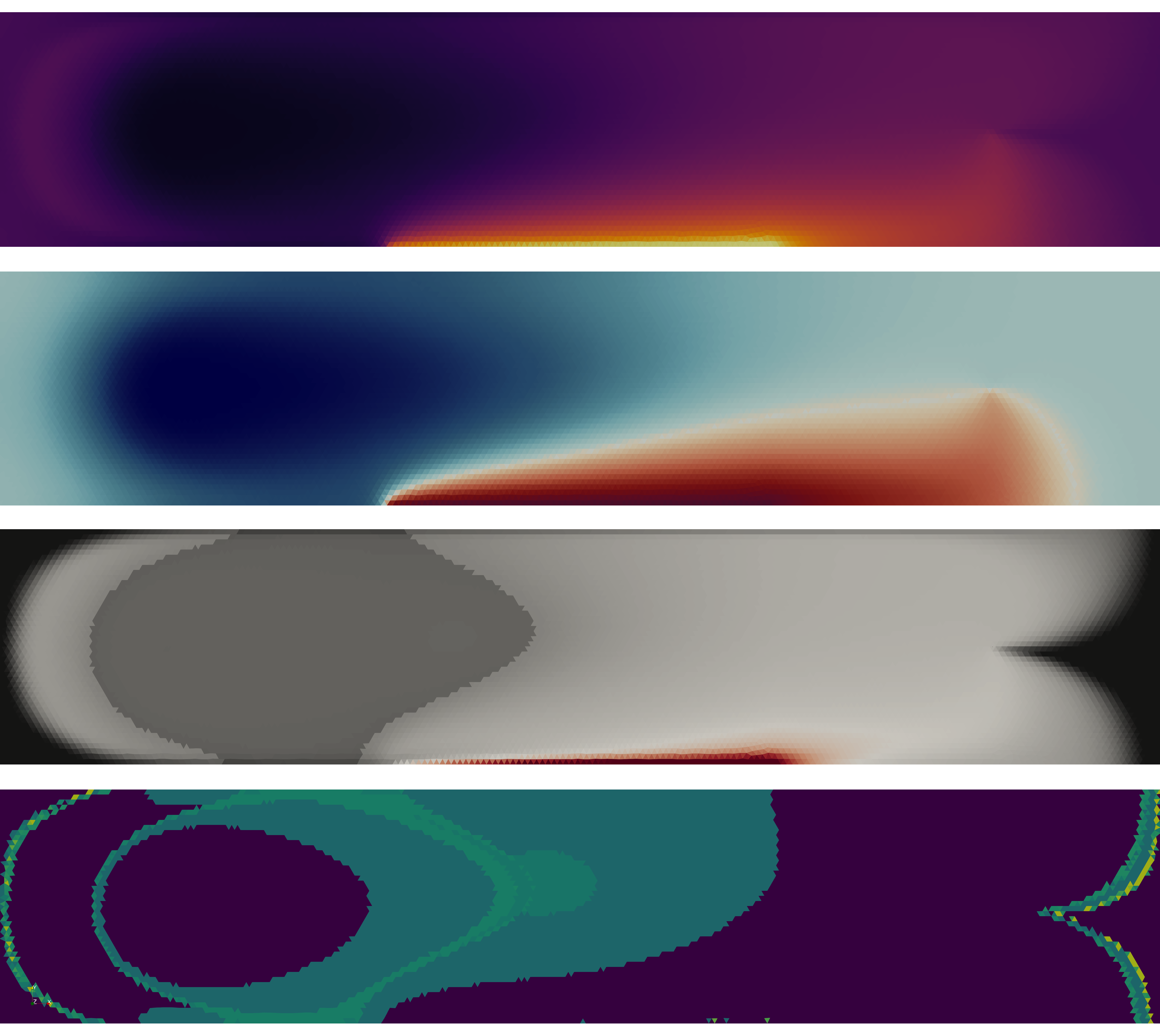}}
    \subfigure[]{\includegraphics[width=0.45\textwidth]{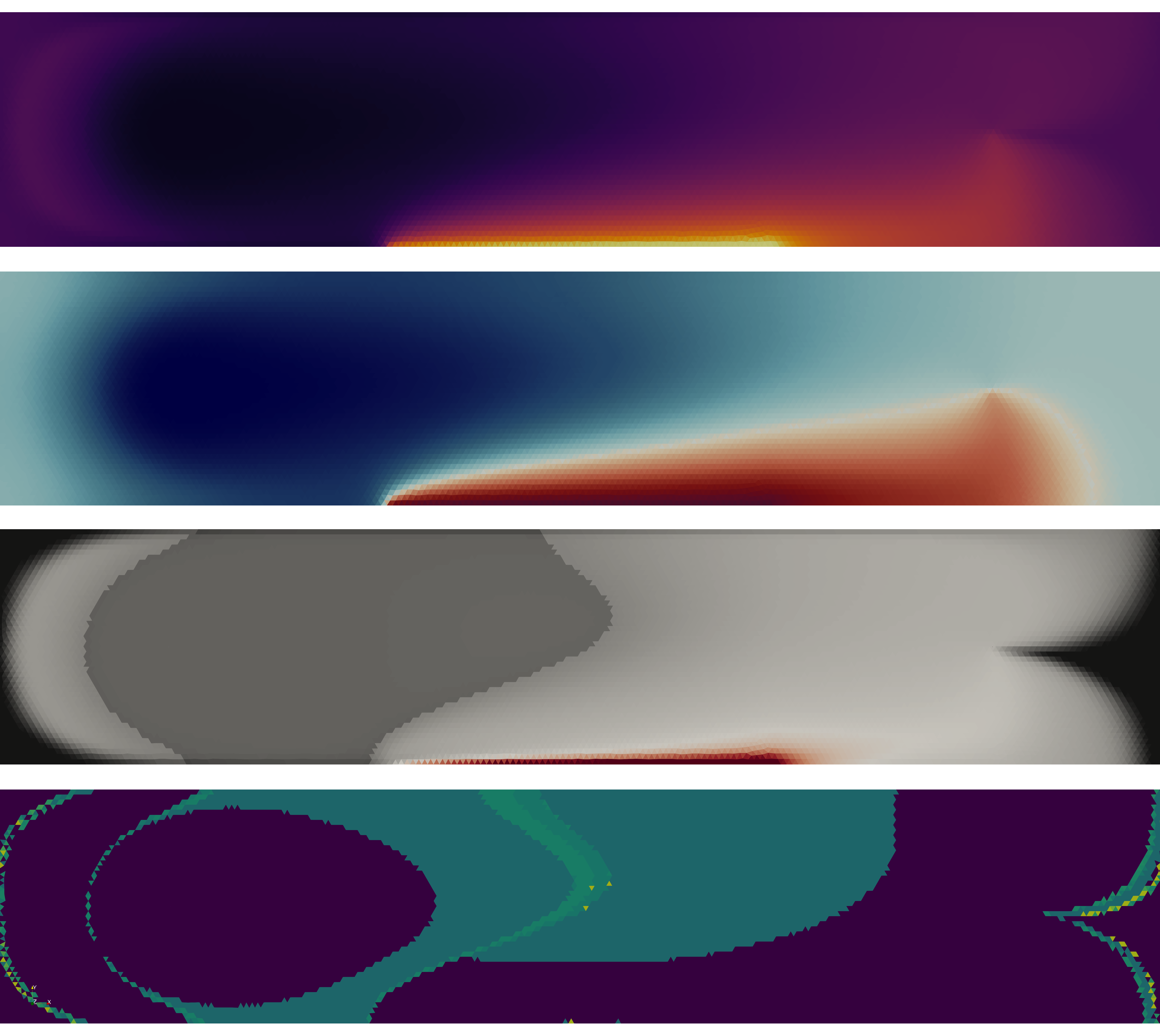}}
    \subfigure[]{\includegraphics[width=0.45\textwidth]{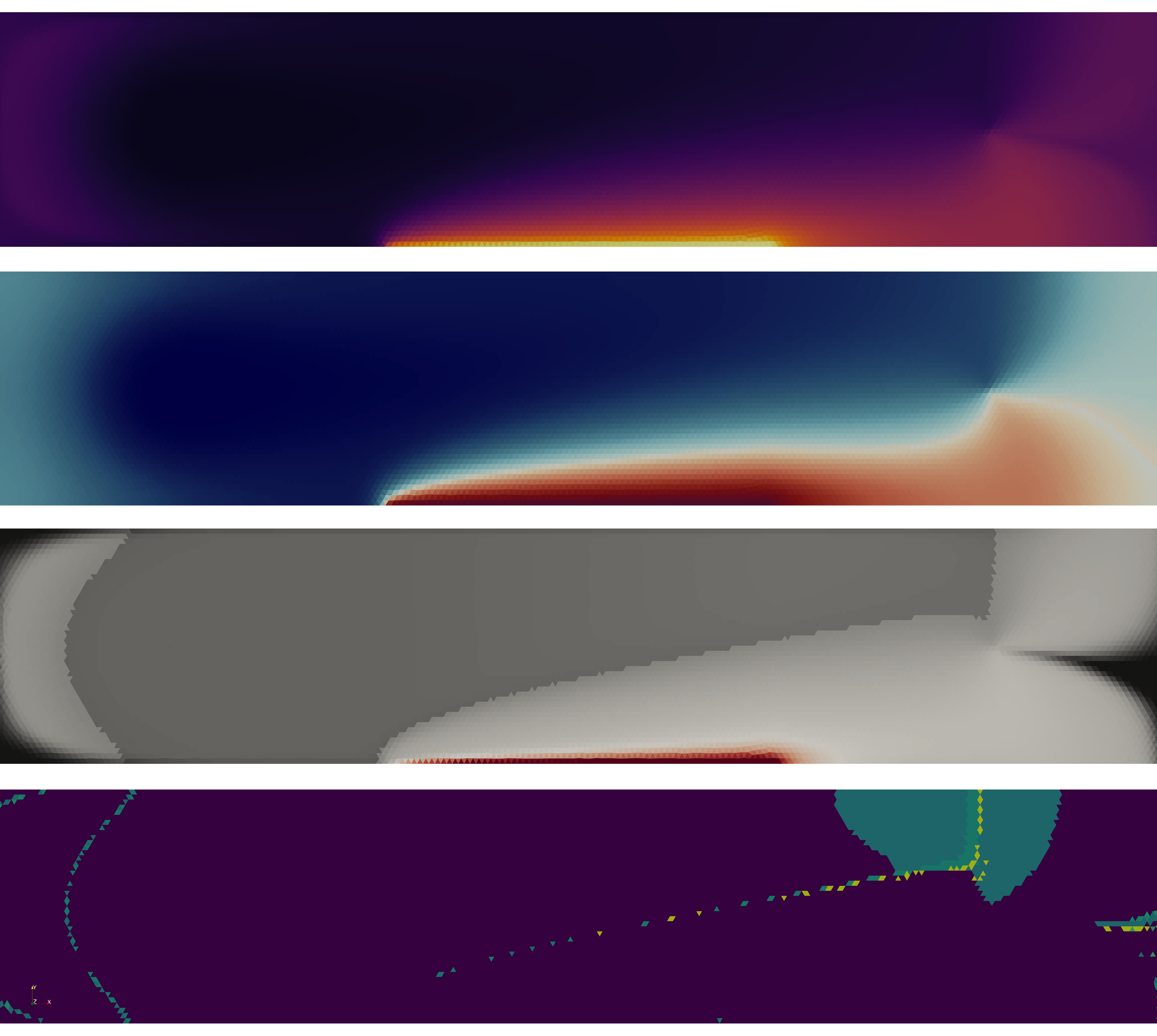}}
    \vspace{-10pt}
    \caption{Simulation after (a) 90, (b) 240 , (c) 300 and (d) 600 days. Rows from top to bottom: specific fluid enthalpy $\specEnthalpy$ [kJ / mol], temperature $T$ [K], gas saturation $s_g$ [-] and local iterations summed over iterations per displayed time step.}
    \label{fig:simulation}
    \vspace{-20pt}
\end{figure}
The domain illustrated in \Cref{figure:simulation-setup} is initially filled with hot, liquid, almost pure water, equilibrated at a pressure of 20 MPa and a temperature of 450 K.
A mixture with 10 \% \cotwo is injected at a constant rate and at 300 Kelvin.
Since transport properties are not state functions, viscosity and thermal conductivity require additional assumptions.
We chose constant values in this section and discuss a more complex model in the following section.
The injection starts sharply, with no delay in time.
As the heat source at the bottom, we chose a temperature of 640 K, surpassing the boiling point of water at the initial pressure.
The simulated time is 600 days.
An initial time step size of $1/2$ day is chosen.
We allow for 10 time step reductions and a minimum time step size of 1 hour.
Otherwise, we consider the simulation unsuccessful.
As the convergence criterion, a global residual tolerance $\epsilon_{g}$ is applied equation-wise.
The domain was discretized with a triangular mesh of mesh size $h_{mesh} = 0.5$ m, resulting in 18466 cells and 184660 degrees of freedom for a two-phase two-component fluid.
The simulation setup is summarized in \Cref{table:pT-ph-setup}.

The solution procedure is configured so that the local equilibrium condition is formulated in terms of pressure and enthalpy, except at the injection point.
At the injection point, the local equilibrium is resolved in terms of $p$ and $T$ since the temperature of the injected fluid is fixed, while $\specEnthalpy$ is updated after the local solver based on the resulting fractions $y_\gamma, x_{\xi\gamma}$.
In the remaining domain, including the production point, $p$ and $\specEnthalpy$ are fixed during the local equilibration and $T$ is updated.
Note, however, that the global system is based on an enthalpy-formulation.
Choosing a local equilibrium calculation in terms of $pT$ or $p\specEnthalpy$ effectively means choosing a part of the global residual which should go to zero through the equilibration.
The equilibration in terms of $p\specEnthalpy$ additionally reduces the residual of the local enthalpy constraint $\Upsilon_{p\specEnthalpy}$ to zero (\Cref{equ:equilibrium-subsystem-h}), compared to the $pT$-equilibration which reduced only the remaining equations in System \eqref{equ:equilibrium-subsystem}.

\vspace{10pt}
\begin{minipage}{\textwidth}
\begin{minipage}[b]{0.4\textwidth}
\centering
\begin{tabular}{l l}
\toprule
domain & $100\times 20$ m \\
$h_{mesh}$ & 0.5 m \\
$K$ & \num{1e-13} m$^2$ \\
$K_l$ & \num{1e-14} m$^2$ \\
$\Phi$ & 0.1 \\
$\rho_s$ & 2950 kg/m$^3$ \\
$h_s(T)$ & $c_s(T - T_i)$ J/kg \\
$c_s$ & 603 J / (kg K) \\
$\kappa_s$ & 1.6736 W/(m K) \\
$\mu_\gamma$ & \num{1e-3} Pa s\\
$\kappa_\gamma$ & 1 W/(m K) \\
$T_h$ & 640 K\\
$r_i$ & $10\times27430$ mol/(m$^3$h)\\
$r_{i, \xi}$ & $(0.9r, 0.1r)$ \\
$T_i$ & 300 K\\
$p_p$ & 19 MPa\\
$p(t_0)$ & 20 MPa\\
$T(t_0)$ & 450 K\\
$z(t_0)$ & $(0.995, 0.005)$ \\
$t_{end}$ & 600 d \\
$\Delta t_0$ & 1/2 d \\
$\epsilon_g$ & \num{1e-7}\\
\bottomrule
\end{tabular}
\captionof{table}{Domain specifications, initial conditions and solver parametrization.}
\label{table:pT-ph-setup}
\end{minipage}
\hspace{15pt}
\begin{minipage}[b]{0.45\textwidth}
\centering
\includegraphics[width=\linewidth]{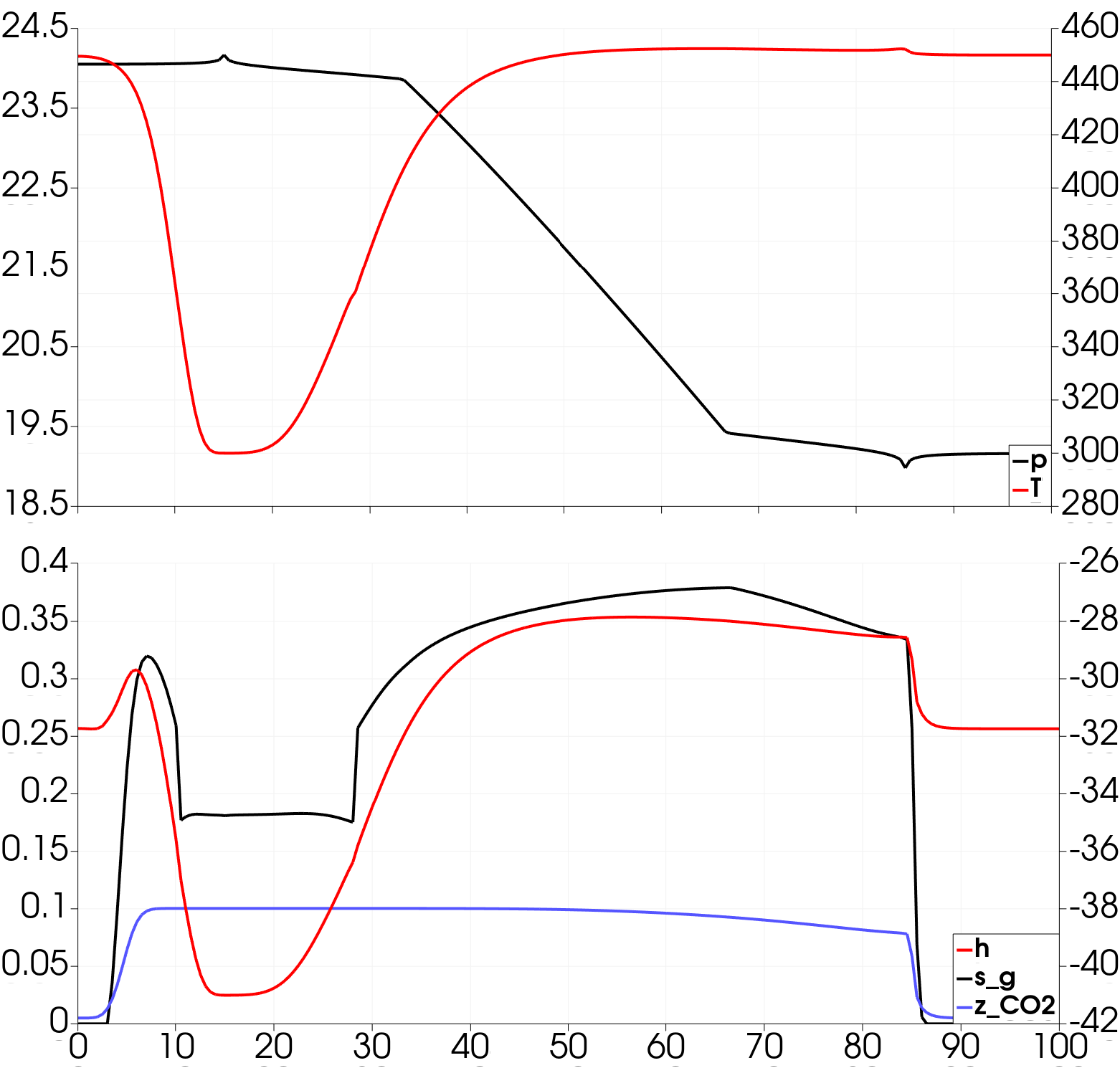}
\captionof{figure}{Top: Pressure $p$ [MPa] (left axis) and temperature $T$ [K] (right axis). Bottom: Gas saturation $s_g$ [-], fraction of carbon dioxide $z_{\text{\cotwo}}$ [-] (left axis) and specific fluid enthalpy $\specEnthalpy$ [kJ / mol] (right axis). The profiles correspond to \Cref{fig:simulation} (a), horizontally through the domain intersecting wells. The injection and production point are positioned at 15 and 85 meters respectively.}
\label{fig:line-plot}
\end{minipage}
\end{minipage}
\vspace{10pt}

\Cref{fig:simulation} shows the results.
We observe two areas where vapor/gas appears.
The first area is a cold two-phase plume, caused by the injected \cotwo which propagates from the injection to the production point.
Under present pressure and temperature conditions, the two-component fluid separates into a gas-like phase with mostly super-critical \cotwo, and a liquid phase with mostly water. 
The second area appears at the heated bottom.
As the temperature exceeds the boiling point of water, a sharp front appears between the fully evaporated boundary layer and the liquid-like water in the domain.
This sharp front reflects the narrow-boiling behavior of water.
We also observe a thin layer of liquid-like water between the evaporated zone at the bottom and the two-phase plume.
The plume bulges over the bottom boundary layer and is asymmetric in the vertical direction.
This is due to the fully evaporated fluid in the bottom boundary and a respective increase in pressure at the bottom.

\Cref{fig:simulation} also shows the local iterations of the phase separation calculations.
The iteration numbers are cell-wise and cumulative over the global iterations which were performed to reach the displayed time steps.
The color bar is scaled logarithmically to reveal how the iteration numbers reflect every profile we observe in the mass and energy variables.
The local solver was idle in large parts of the domain, where the state of the fluid in terms of mass and energy did not change.
With the equilibrium problem being an optimization problem and our choice of using the previous global iterate state as the initial guess, we avoid phase separation calculations where they are not needed.
In \Cref{fig:simulation} (a) we see that the local solver is mainly active at the three observed fronts: the cold front emanating from the injection well, the two-phase plume reaching the production well, and the evaporating bottom layer; i.e., where either mass or energy or both change.
In \Cref{fig:simulation} (b) - (d) the large mixing areas of hot and cold fluid are shown from the perspective of the local solver.

\begin{wrapfigure}{r}{0.48\textwidth}
    \vspace{-10pt}
    \centering
    \includegraphics[height=4cm,keepaspectratio ]{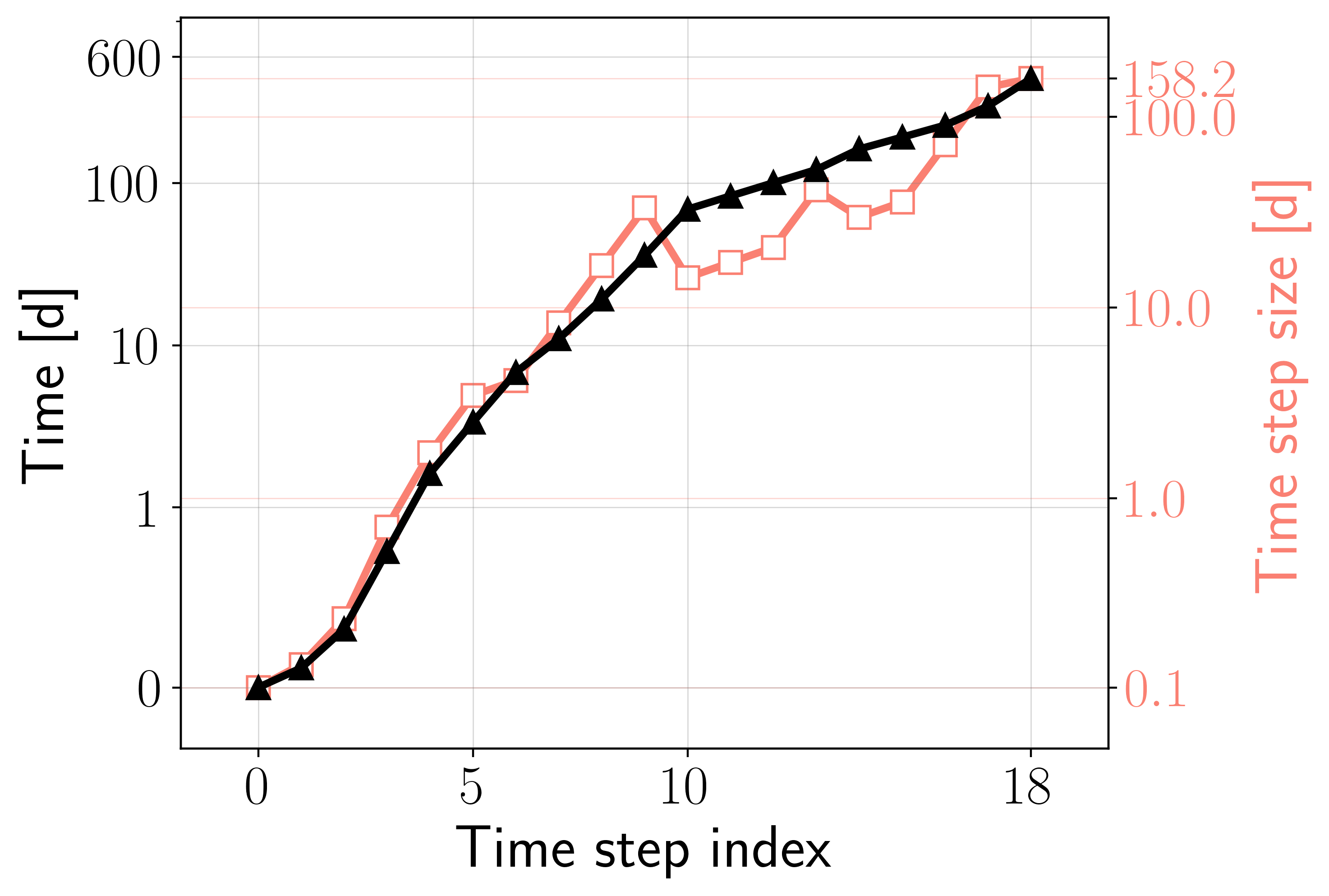}
    \vspace{-10pt}
    \caption{Progress in time  (left axis) and time steps size (right axis) in days.}
    \label{fig:time-progress-ph}
    \vspace{-10pt}
\end{wrapfigure}
The cold two-phase plume shows another effect around the injection point.
The saturation profile displays a ring, marking a steep drop in gas saturation after an initial increase at the beginning of the injection.
Similarly, we observe a wave of higher enthalpy being transported away from the injection.
\Cref{fig:line-plot} plots the profiles along a horizontal line between the two wells.
As injection begins, pressure increases and mass enters an initially hot rock, leading to increased evaporation and fluid enthalpy.
This enthalpy is transported away from the injection well, and as cold fluid enters the domain, the enthalpy drops again.
That is, under hot conditions in the domain, a sudden and high injection rate can cause more fluid around the well to transition into the gas phase before condensation sets in due to the cold mass entering the domain.
The resolution of this wave is also visible in the local iteration numbers, with higher activity where the enthalpy rises and where it falls again.

\Cref{fig:time-progress-ph} shows the progress in time of the $p\specEnthalpy$-based simulation.
A total of 18 time steps were required to reach the end time of 600 days.
The time step size continuously increases as we progress in time.
Around 90 days into the simulation, the algorithm was forced into reducing the time step size.
Comparing with \Cref{fig:simulation}, this is the time when the two-phase plume meets the fully evaporated bottom boundary layer.
Other reductions in time step size are observed in the days after, which is the mixing period where the cold plume passes above the bottom layer and the hot and cold areas start exchanging heat.
The last time step was computed with $\Delta t = 158.2$ days, after the dynamic period is passed and the cold \cotwo breaks through.

\begin{figure}[h]
  \centering
    \includegraphics[height=4.6cm, keepaspectratio]{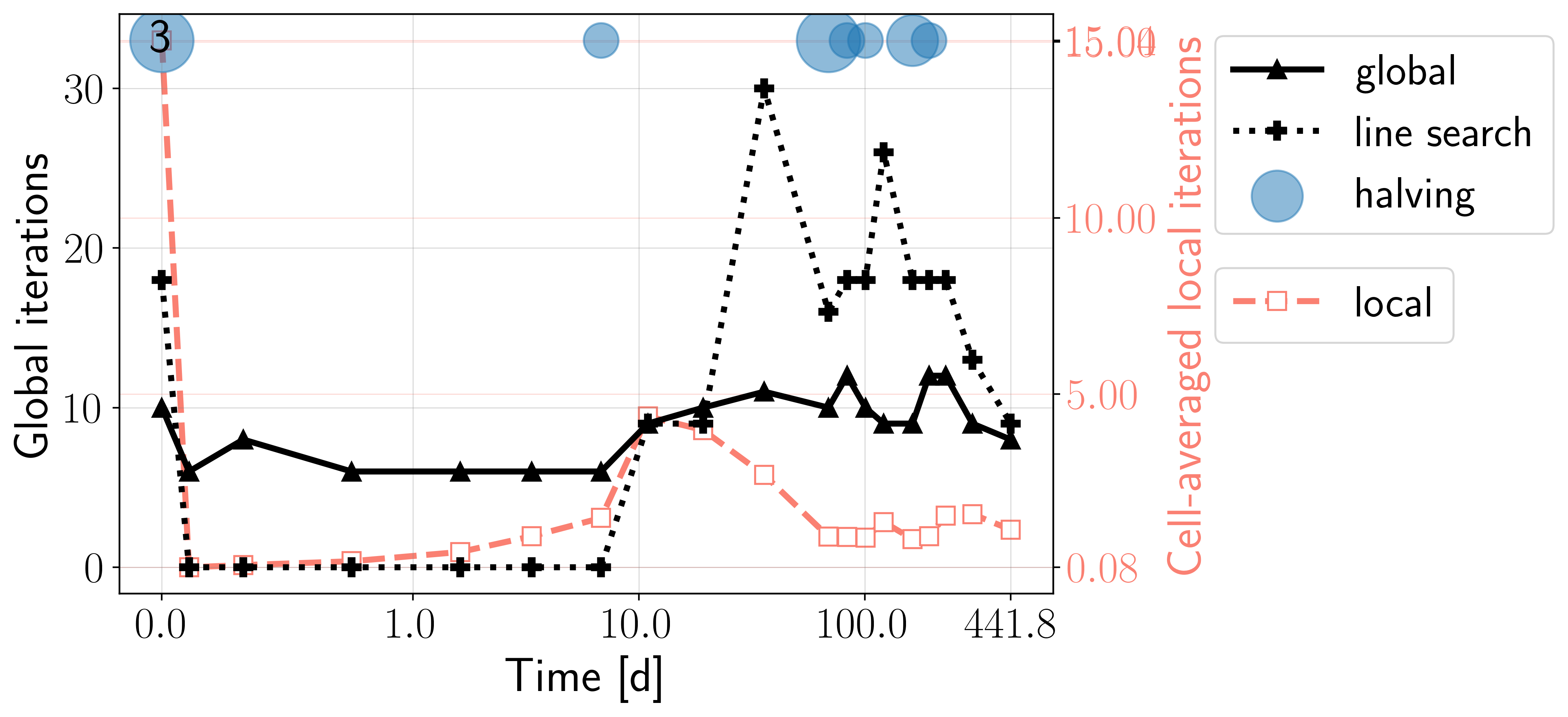}
    \vspace{-10pt}
      \caption{Global Newton and line search iterations (left axis) and local iterations (right axis) of the equilibrium solver summed over the iterations per time step and averaged by the number of cells.
      The number of performed time step reductions is indicated with blue circles and their size. Most reductions were performed in the period of hot-and-cold mixing.}
  \label{fig:iteration-numbers-ph}
\end{figure}
\Cref{fig:iteration-numbers-ph} shows the number of iterations per time step and the time step reductions.
On average, the algorithm required 9 global Newton iterations to converge.
The local solver had two periods of intense calculations.
The first was at the beginning, which required a resolution of the equilibrium state as the pressure between the wells was established.
This transient phase also required a reduction of the time step size, from $\Delta t_0 = 1/2$ day to approximately 4 hours.
The second was after 20 days when the first cells at the bottom started to evaporate.
The line search algorithm was most required in the period after the 90th day, when the cold and boiling fluid started mixing.
With the exception of these events, many cells, especially in the upper half of the domain, required little to no computational resources, due to the local problem being an optimization problem, and exiting if the previous state is close enough.
\subsection{Enhanced model with Lohrenz-Bray-Clark viscosity correlations}
\label{subsection:lbc}
\begin{figure}
    \centering
    \subfigure[]{\includegraphics[width=0.24\textwidth]{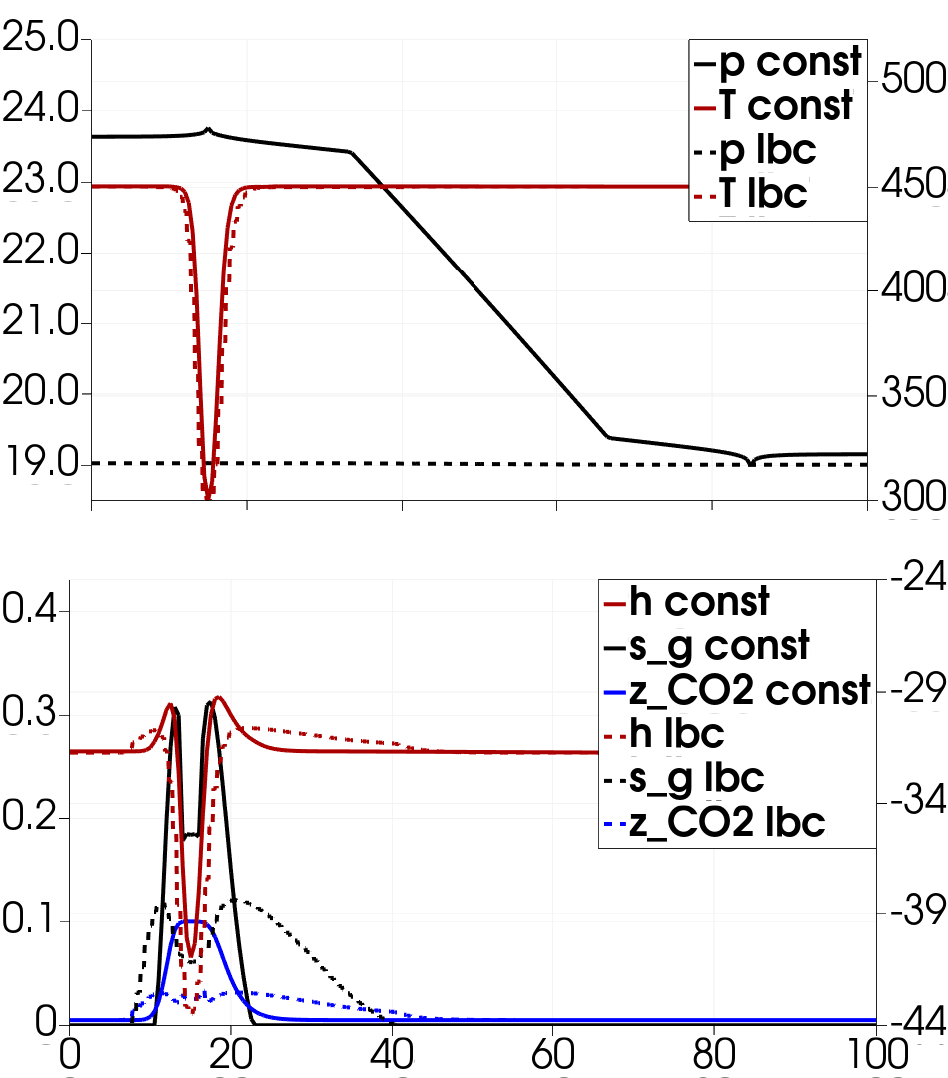}}
    \subfigure[]{\includegraphics[width=0.24\textwidth]{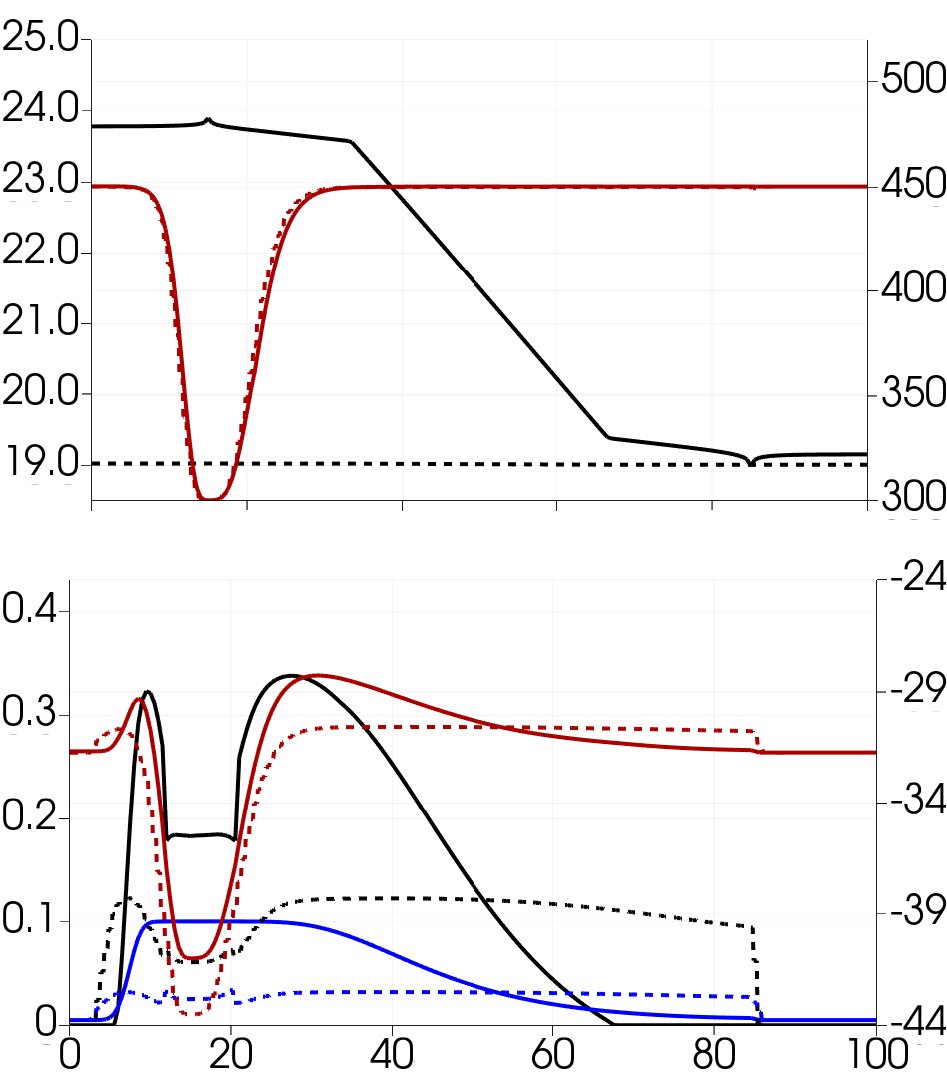}}
    \subfigure[]{\includegraphics[width=0.24\textwidth]{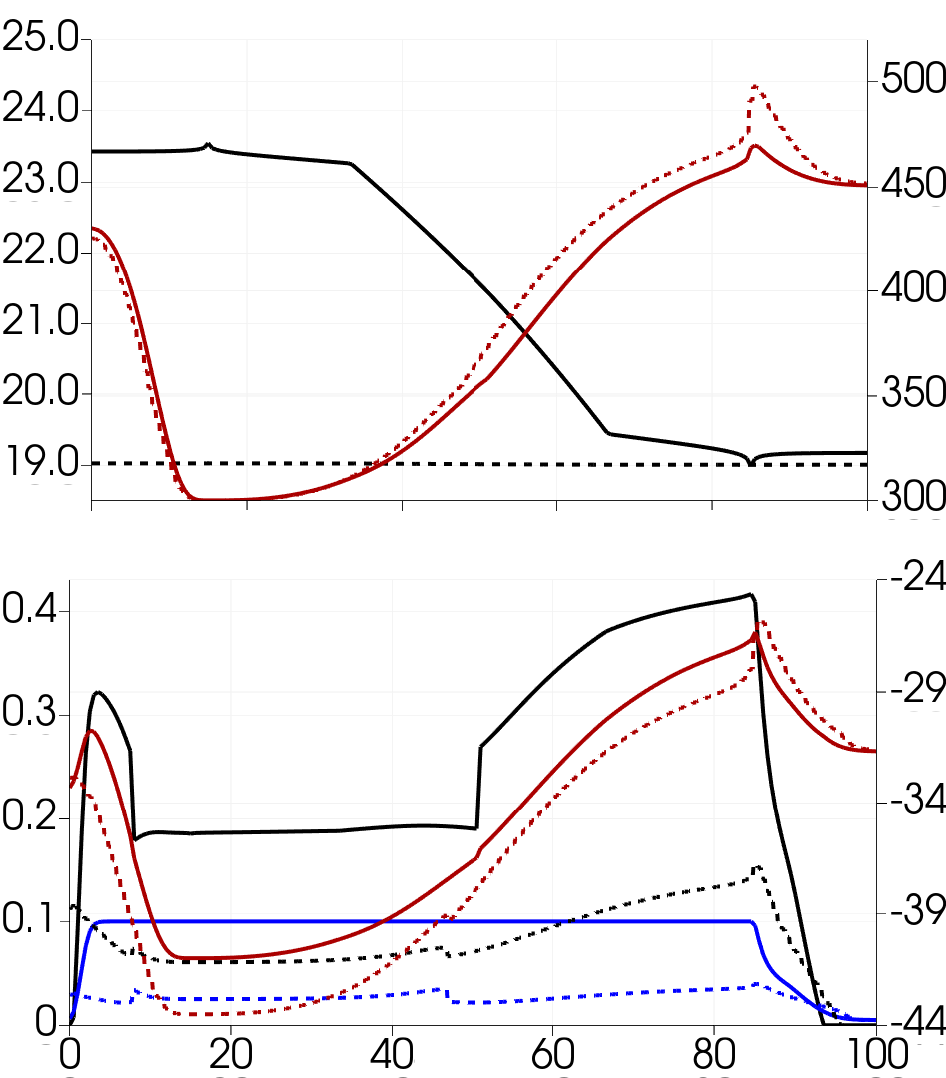}}
    \subfigure[]{\includegraphics[width=0.24\textwidth]{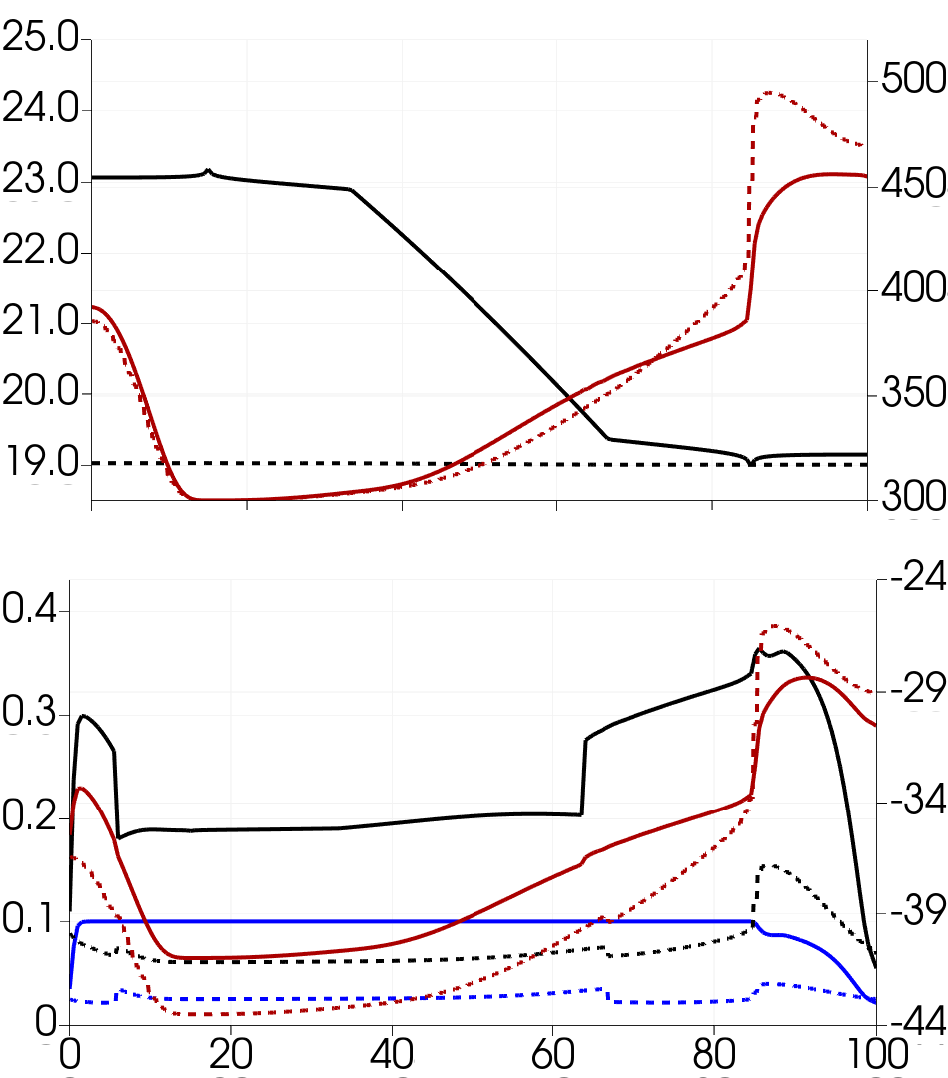}}
    \vspace{-10pt}
    \caption{Simulation with LBC correlations after (a) 3, (b) 30 , (c) 300 and (d) 600 days. Profiles along the horizontal line connecting injection and production well are shown. Displayed are pressure [MPa] and temperature [K] on the top left and right axis respectively, \cotwo fraction and gas saturation [-] on the bottom left axis, and specific fluid enthalpy [kJ / mol] on the bottom right axis. The simulation using the LBC correlations is represented by dashed lines.}
    \label{fig:lbc-simulation}
\end{figure}

\begin{wrapfigure}{r}{0.4\textwidth}
\vspace{-10pt}
  \centering
    \includegraphics[width=\linewidth]{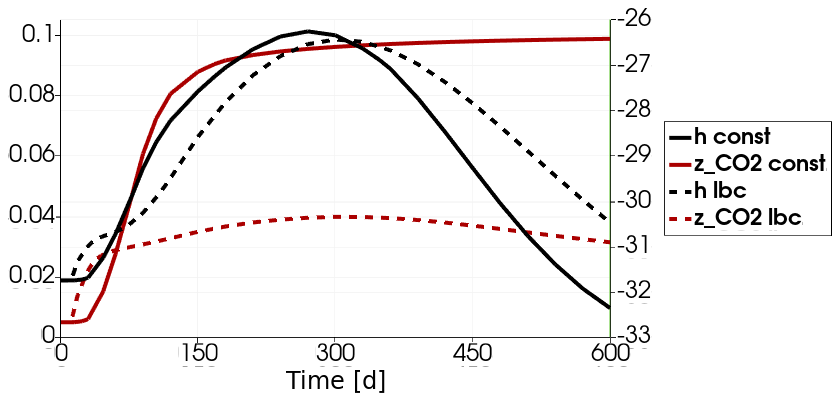}
    \vspace{-10pt}
  \caption{Fraction of \cotwo (left axis) and specific fluid enthalpy $\specEnthalpy$ [kJ / mol] (right axis) obtained with the LBC correlations. The values are plotted over time in the production well.}
  \vspace{-10pt}
  \label{fig:lbc-production}
\end{wrapfigure}
The setup discussed in \Cref{subsection:ph-simulation} is dominated by advection.
For a fluid mixture with two components and 2 phases with different physical states, the constant viscosity assumption is a crude simplification.
In this section, we additionally apply the Lohrenz-Bray-Clark model for viscosity which is a commonly used correlation in, e.g., petroleum and CCS applications \cite{gharbia2019,li2023}.
This motivates a test of the presented mathematical model and solution strategy using this enhanced fluid model.
Since the example is dominated by advection, we keep the constant values for thermal conductivity but chose different values for the phases.
The liquid-like phase is assigned a value of $\kappa_l = 0.6$, and the gas-like phase a value of $\kappa_g = 0.06$ [W/ m K].

\Cref{fig:lbc-simulation} compares the profiles of the primary variables of interest along a horizontal line that connects the two wells.
We first notice that the pressure in the domain after 3 days is almost flat at 19 MPa, corresponding to the value at the production well.
This is caused by the LBC viscosity taking a significantly lower value than the original constant value of \num{1e-3} Pa s, leading to a rapid depletion of the domain.
The \cotwo and respectively the two-phase plume propagate much faster through the domains as is evident on \Cref{fig:lbc-simulation} (a)-(b).
Furthermore, due to the fast propagation of the gas-like phase, which is where most of the \cotwo resides, the \cotwo concentration never accumulates to reach the value of 0.1 (the mass ratio in the injection rate) in any cell.
This rapid passing through also cools the domain less, as is evident in the slightly higher temperature profile in \Cref{fig:lbc-simulation} (c).
However, the specific fluid enthalpy is lower, which we attribute to the smaller amount of \cotwo in the domain and hence to a reduction in the gas phase saturation and thermodynamic work.

\Cref{fig:lbc-production} shows the estimated production curves for specific fluid enthalpy and \cotwo fraction obtained with the two models for viscosity. Both models lead to an increase in specific enthalpy as the injected fluid reaches the production well, with the LBC model predicting a slower increase, a lower peak value, but also a slower decrease after the peak:
The simple viscosity estimates though show a fall of produced energy to pre-injection levels already after 500 days, while the LBC case still displays higher energy levels after 600 days.
The produced fraction of \cotwo are significantly lower with LBC, with a peak concentration of below 0.04, while the constant viscosity gives a gradual increase towards the injected concentration of 0.1.

\begin{figure}[h]
  \centering
    \includegraphics[height=4.6cm, keepaspectratio]{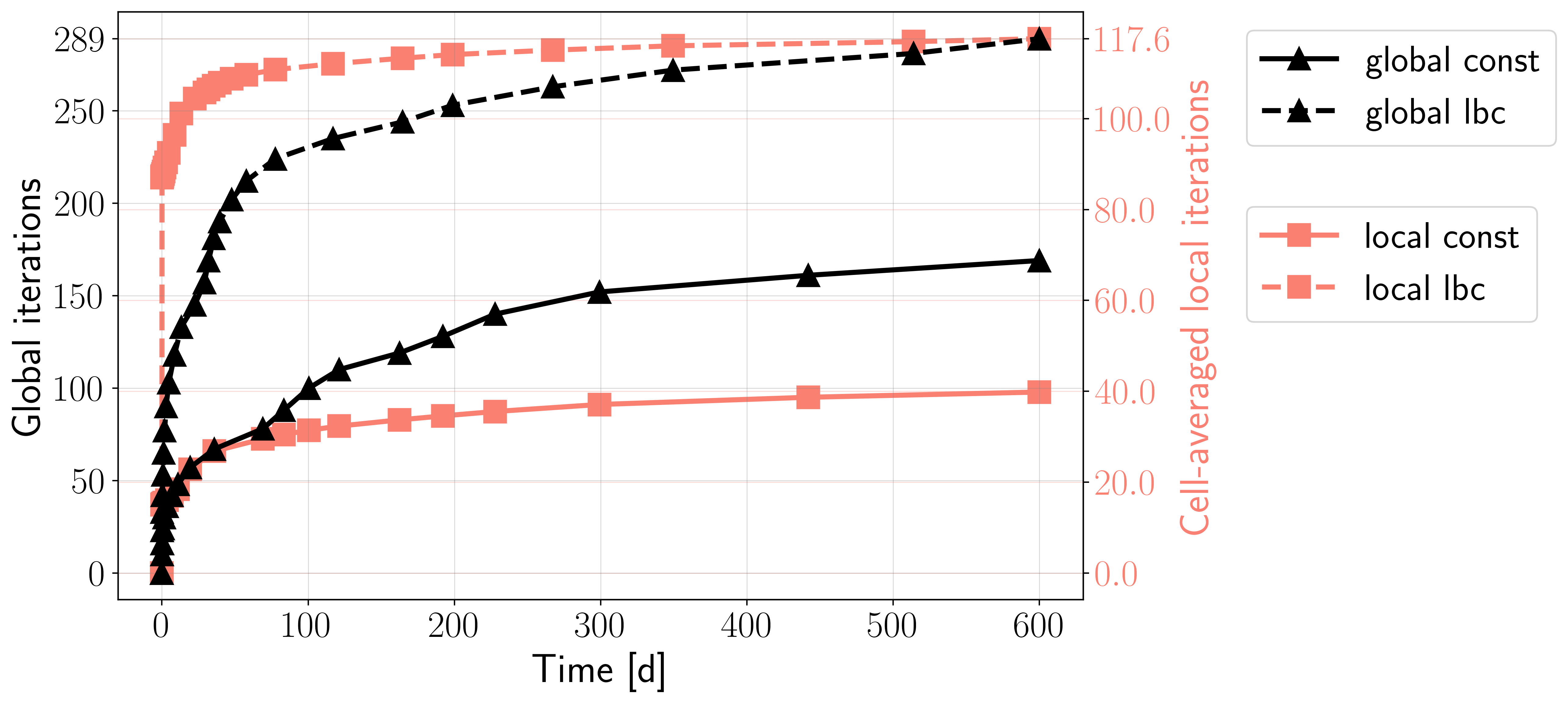}
    \vspace{-10pt}
  \caption{Cumulative numbers of iterations for cases with constant viscosity and LBC-viscosity. The local iterations are cell-averaged values.}
  \label{fig:lbc-comparison-iter}
\end{figure}
\Cref{fig:lbc-comparison-iter} compares the cumulative numbers of global and local iterations for the constant- and LBC-viscosity models.
The local iteration numbers represent the cell-averaged values summed over iterations per time step.
Using the LBC viscosity model, the average cell has undergone 117.6 local iterations after 600 days of simulation, while only 39.8 for the constant-viscosity case.
Analogously, the more complex fluid model required a total of 289 global nonlinear iterations, while the other finished after 169 iterations.
Most of the additional iterations in the LBC model are spent in the beginning during the transient phases of domain depletion and boiling in the bottom layer.
These processes change the pressure and enthalpy conditions in the domain significantly, leading to strongly varying phase mobilities and hence more iterations.
\subsection{Impact of local solver configurations on number of nonlinear iterations}\label{subsection:comparison}
\Cref{subsection:ph-simulation} demonstrated a robust solution strategy, where the isenthalpic equilibrium conditions applied in the interior of the domain were able to capture the non-isothermal dynamics.
We will now address the efficiency and necessity of applying the isenthalpic equilibrium calculations.

\Cref{subsection:phase-equilibrium} shows that the equilibrium formulation based on $pT$ is a subset of the $p\specEnthalpy$-formulation in terms of equations and variables.
Both are included in the mathematical model presented and usable when required.
In order to verify the necessity of the isenthalpic approach, we conducted the simulation presented for both $pT$ and $p\specEnthalpy$, for a series of grid refinements, and for a time period of 6 months.
We recall that an isothermal equilibration step does not reduce the residual of \Cref{equ:equilibrium-subsystem-h}, as opposed to an isenthalpic equilibration.
The global system is in both cases the same, i.e enthalpy based.

\begin{figure}[h]
  \centering
    \includegraphics[height=4.6cm, keepaspectratio]{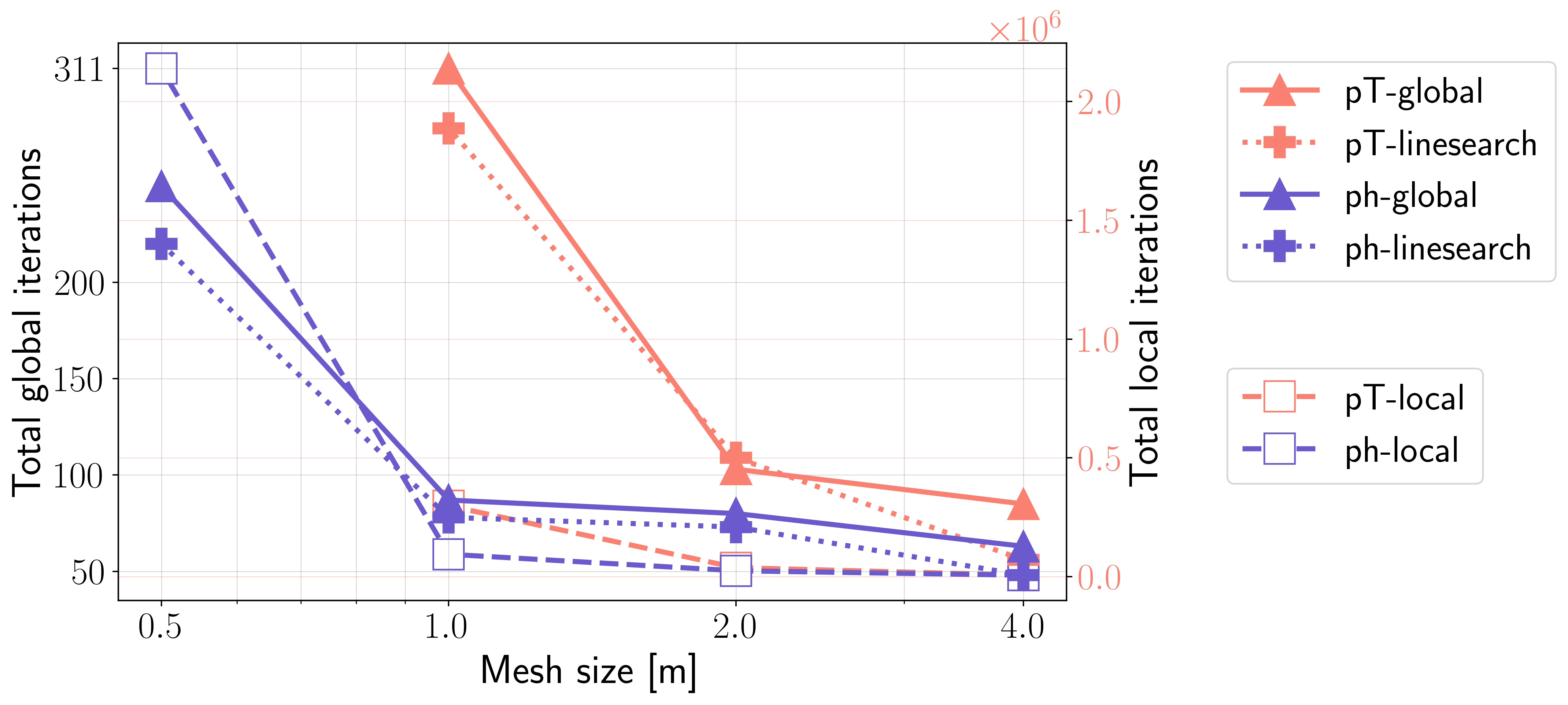}
    \vspace{-10pt}
  \caption{Total number of global Newton and line search iterations (right) and total number of local iterations (left) of the equilibrium solver per refinement. The simulation using a local equilibration in terms of $pT$ failed for the highest refinement.}
  \label{fig:iteration-per-refinement}
\end{figure}

\Cref{fig:iteration-per-refinement} shows the number of iterations per grid refinement for both approaches.
After the third refinement, the $pT$-approach was unable to complete the simulation successfully.
It failed to converge with the minimal admissible time step size of 1 hour.
This is expected since at $h_{mesh}=0.5$ the grid is fine enough to display the fully evaporated layer of water at the bottom for which isothermal calculations are not suitable.
Naturally, allowing the time step size to decrease further could result in success, since the global formulation includes a local enthalpy constraint \eqref{equ:equilibrium-subsystem-h}.
But this is not desirable.
Moreover, \Cref{fig:iteration-per-refinement} shows that the use of isenthalpic equilibration consistently leads to fewer iterations, globally and locally.
That is, solving a larger subsystem within global iterations accelerates global convergence more than solving a smaller system.
In the presented case, the $p\specEnthalpy$-subsystem is larger by only 1 local degree of freedom.
But because of the setting and the narrow-boiling physics, this becomes significant.

\begin{table}
\centering
\begin{tabular}{@{}l | l l |l l | l l | l@{}}
\toprule
\makecell[l]{Equilibrium\\($h_{mesh}$)} & $p\specEnthalpy$(4) & $pT$(4) & $p\specEnthalpy$(2) & $pT$(2) & $p\specEnthalpy$(1) & $pT$(1) & $p\specEnthalpy$(0.5) \\
\midrule
\makecell[l]{linear solve\\time} & \makecell[l]{0.0064\\ (0.40)}& \makecell[l]{0.0065\\ (0.55)}&  \makecell[l]{0.0192\\ (1.54)}& \makecell[l]{0.0210\\ (2.08)}& \makecell[l]{0.0764\\ (6.65)}& \makecell[l]{0.0776\\ (15.20)}& \makecell[l]{0.4030\\ (68.11)}\\
\addlinespace
\makecell[l]{local solve\\time} & \makecell[l]{0.0286\\ (2.12)}& \makecell[l]{0.0298\\ (2.86)}& \makecell[l]{0.0751\\ (6.84) }& \makecell[l]{0.0949\\ (10.72)}& \makecell[l]{0.1391\\ (13.64)}& \makecell[l]{0.1626\\ (36.74)}& \makecell[l]{0.2930\\ (58.60)}\\
\addlinespace
\makecell[l]{No. time\\steps} & 10 (0) & 10 (0) & 10 (0) & 11 (2) & 10 (0) & 18 (12) & 18 (12) \\
\addlinespace
\makecell[l]{No. global\\ iterations} & 63 (0) & 85 (0) & 80 (0) & 99 (4) & 87 (0) & 195 (116) & 169 (81) \\
\addlinespace
\makecell[l]{T. No. local\\iterations} & 6533 & 6979 & 25121 & 38167 & 94556 & 298742 & 2138763 \\
\bottomrule
\end{tabular}
\caption{\label{table:metrics} Comparison of solver metrics for both equilibrium conditions per refinement. Times are given in seconds, as pairs of average and total value. Numbers of time steps and iterations include the number of wasted executions in brackets.
Local iterations are given as a total number.}
\end{table}

\Cref{table:metrics} summarizes the solver metrics for all refinements and the two equilibrium specifications.
Local equilibration based on $pT$ is worse on all metrics.
The fact that it results in the solution procedure spending more overall time in the local solver, even though the system is smaller, is explained by the layer of cells near the bottom boundary, which is challenging to equilibrate.
The local solver, which is a minimization algorithm, starts with the current iterate state of $\varY$ as an initial guess.
If that fails or the maximum number of local iterations is reached, a new attempt is made at equilibrating the fluid with an in-built initial guess procedure.
Therefore, the $pT$-condition results in the local solver performing more computations in individual cells in this layer and significantly increasing the overall time.
Although we expect the isothermal local equilibration to perform worse than the isenthalpic one, it is surprising to see that the isothermal equilibration also has a deteriorating impact on the total time spent in the linear solver.
We know that in the narrow-boiling case enthalpy is sensitive to temperature, and reciprocally, temperature is insensitive to enthalpy.
It appears that an insufficiently precise equilibration; i.e., a mismatching $T-\specEnthalpy$ pair, affects the inverse in the Schur complement, which effectively leads to a representation $T(\specEnthalpy)$ on the discrete level.
This can contribute to the ill-conditioning of the matrix, and hence influence processes within the direct solver.
The presented solution strategy leaves room for optimization by choosing the equilibrium conditions in cells individually (which we did at the injection point), but we conclude that this is not necessary for this example and stick to the isenthalpic local calculations.

As a final investigation, we want to answer the question of whether it is required to equilibrate the fluid at all, how often, and up to which precision.
The equilibrium solver is per se an encapsulated part of the algorithm, which can be invoked at various steps of the global solution procedure.
According to \Cref{subsection:nonlinear-solver}, we perform the equilibration here on every cell after a global Newton iteration.
To reduce the amount spent on the local solver, since equilibration is not always required, we introduce a global iteration stride, performing the equilibration only every couple of iterations.
We conducted a parameter study, running the simulation using $p\specEnthalpy$-conditions for 6 months and analyzing the impact of performing the equilibration every $n\in\{1,\dots,5\}$ global iteration and not at all.
We also investigated the impact of local solver tolerance on the overall efficiency of the solution strategy.

\begin{figure}[h]
\centering
\begin{minipage}[t]{0.45\textwidth}
    \centering
    \includegraphics[width=\linewidth]{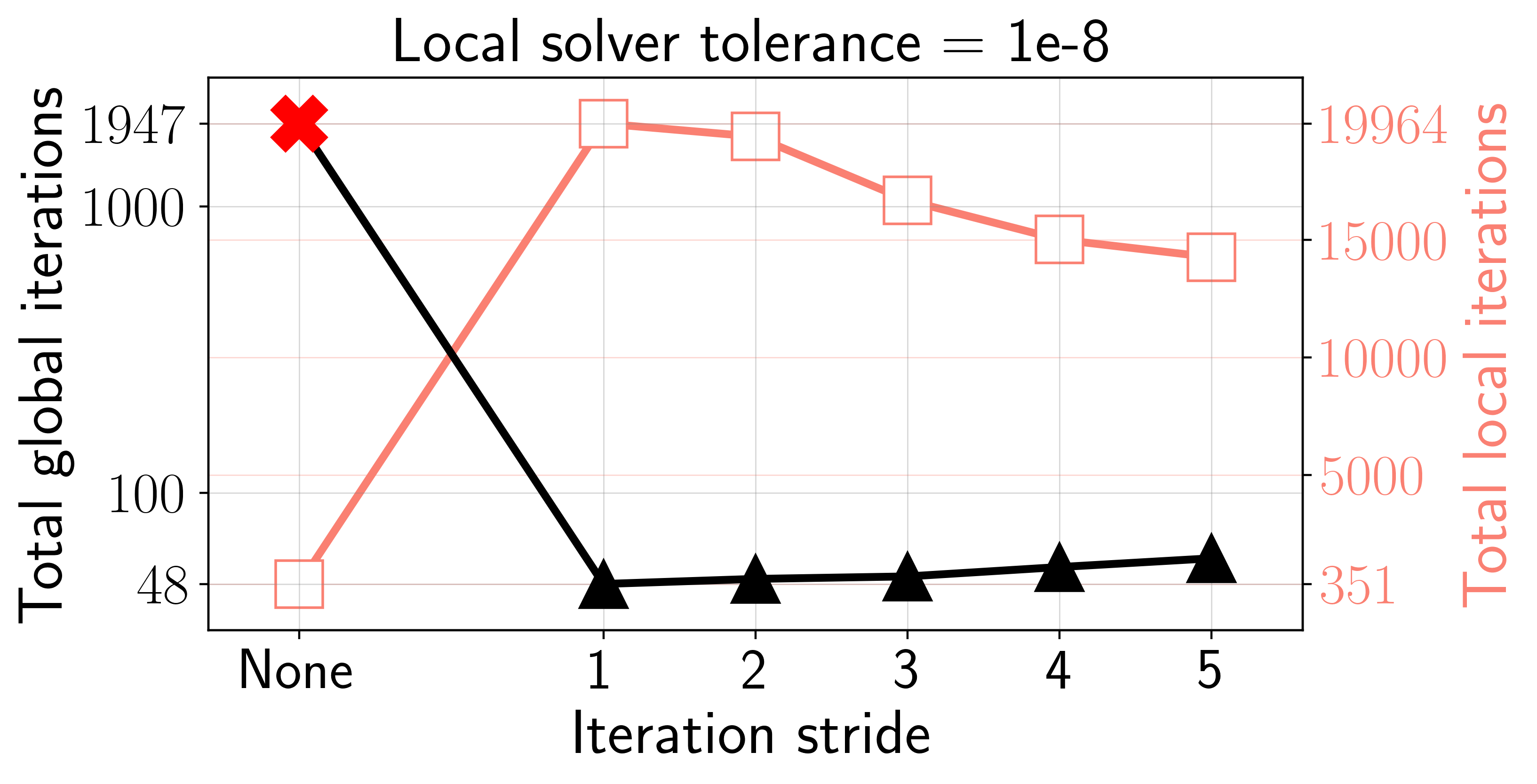}
    \vspace{-10pt}
    \caption{The total number of iterations depending on the local tolerance.
    Performing the equilibration every fifth iteration leads to 62, every iteration to a total of 48 global iterations.
    The local solver is called during the computation of initial values for secondary variables $\varY$, hence the number of local iterations is non-zero everywhere.}
    \label{fig:iter-per-stride}
\end{minipage}
\hfill
\begin{minipage}[t]{0.45\textwidth}
    \centering
    \includegraphics[width=\linewidth]{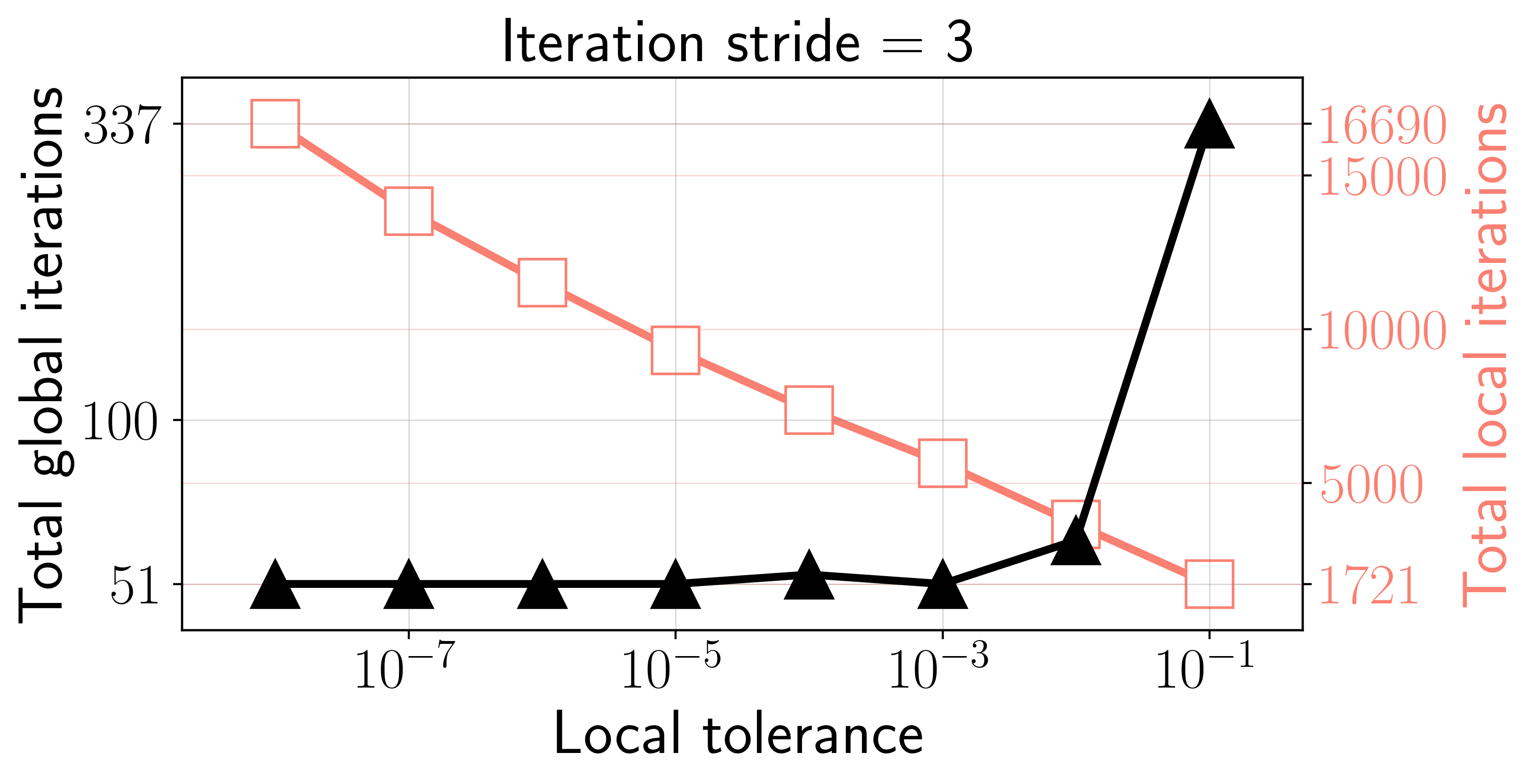}
    \vspace{-10pt}
    \caption{The total number of iterations based on the chosen stride.
    Below the local tolerance of $\num{1e-3}$ the number of global iterations settles around 51.}
    \label{fig:iter-per-ftol}
\end{minipage}
\end{figure}
\Cref{fig:iter-per-stride} shows, that equilibrating the fluid every global iteration compared to doing it every fifth reduces the total number of global iterations from 62 to 48, an improvement of 23 \%.
Not calling the local solver at all causes the required number of global iterations to increase drastically, and in this setup to exit the simulation unsuccessfully.
Reciprocally, performing the equilibration every fifth iteration leads to a decrease in local iterations of approximately $1/4$.
Taking parallelization of the local problem into account, this has significant implications for large-scale simulations supported by suitable hardware.

\Cref{fig:iter-per-ftol} shows the number of global and local iterations depending on the chosen local tolerance $\epsilon_1$ for the equilibrium solver \eqref{equ:local-update-Y}.
On the one hand, omitting to solve the local problem with a sufficient precision has an equally deteriorating effect on the number of global iterations as not solving it at all.
On the other hand, reducing the local tolerance below $\epsilon_1 = \num{1e-3}$ does not show a significant improvement, while, naturally, increasing the number of local iterations, since they scale linearly with the order of the local tolerance.
The local tolerance of $\epsilon_1 = \num{1e-3}$ turns out to be a beneficial choice in terms of both global and local iterations.

Finally, we disclose that for the results presented in \Crefrange{fig:simulation}{fig:iteration-per-refinement}, the nonlinear solver was parameterized to execute the local equilibration every third global Newton iteration with a local tolerance of $\epsilon_1=\num{1e-3}$.
Although it is necessary to perform the equilibration in order to converge at all, executing the local solver every iteration can conflict with the overall progress of the Newton method.
This is mostly the case when a sharp front, such as the narrow-boiling fluid at the heated boundary, propagates slowly, leading to oscillations between gas and liquid states in between global Newton updates and local equilibration.
That is, the correction of the Newton update in the form of an equilibrium calculation must be used with a proper knowledge of the expected dynamics.
Choosing to perform the equilibration every third iteration, we give priority to the global Newton update.
It is the latter that reduces the whole residual $\sysF$ leading to convergence, while local equilibration solves only parts of the problem.

\begin{remark}
    We acknowledge a quantitatively small dependency of the result on the hardware and the operating system.
    The simulations presented here were performed on a Linux platform (Debian GNU/Linux 13 trixie) using an AMD x86-64 CPU.
    The provided docker container eliminates the OS-factor.

    We suspect that the usage of just-in-time compilation of parts of the code in combination with direct linear solvers for ill-conditioned systems is the reason why attempts to reproduce the exact numbers across platforms fail.
\end{remark}

\WFfill
\WFclear
\FloatBarrier
\section{Conclusion}
Using the persistent-variable formulation presents a mathematical approach to continuously express the coupling between thermal compositional flow and the thermodynamic problem of phase separation.
It introduces phase separation as a closed subproblem in the overall mathematical model.
By extending this approach to an isenthalpic local equilibrium formulation, we are able to tackle challenging high-enthalpy scenarios and narrow-boiling phenomena.
We demonstrated that embedding a local thermodynamic solver within a global Newton solver significantly enhances the simulation, allowing us to tackle the nonlinearities arising from equation-of-state-based fluid models.
Since the persistent-variable formulation states phase separation as a closed subproblem, the local solver effectively reduces the global residual by solving the subproblem between global iterations.
It opens not only the possibility to construct nonlinear solvers, but also for numerical analysis.
Our approach is - in a sense -  a splitting-type approach for problems with local-global interactions.
By resolving the instant-time scale of local equilibrium, we achieve faster and robust convergence with fewer global nonlinear iterations compared to the traditional monolithic approach.
Furthermore, the flexibility of the framework - using a consistent and modular set of equations for isothermal and isenthalpic equilibrium calculations — streamlines the modeling of complex phase behavior and offers a scalable approach to simulating thermal compositional flow.
One remaining question is how to optimally use the modularity of the persistent-variable model to perform fluid equilibration locally under different specifications or to include the equilibration in terms of more complex state definitions such as isochoric ones.
That is, a suitable cell-wise criterion that decides when to use $pT$, $p\specEnthalpy$ or other conditions, could further increase the overall performance.

\section*{CRediT authorship contribution statement}

\noindent\textbf{Veljko Lipovac:} Conceptualization, Methodology, Software, Writing - Original Draft, Visualization.
\\
\textbf{Omar Duran:} Conceptualization, Methodology, Software, Validation, Supervision, Writing - Review \& Editing.
\\
\textbf{Eirik Keilegavlen:} Conceptualization, Supervision, Writing - Review \& Editing.
\\
\textbf{Inga Berre:} Conceptualization, Supervision, Writing - Review \& Editing, Project administration, Funding acquisition.

\section*{Declaration of Competing Interest}
\noindent The authors declare that they have no known competing financial interests or personal relationships that could have appeared to influence the work reported in this paper.

\section*{Acknowledgments}
\noindent This result is part of a project has received funding from the European Research Council (ERC) under the European Union’s Horizon 2020 research and innovation program (grant agreement No 101002507).

\section*{Data availability}
\noindent The data and source code for the results presented herein is available, and the plots can be reproduced using a Docker container available at:\newline
\url{https://doi.org/10.5281/zenodo.17335273}.
\FloatBarrier
\allowdisplaybreaks
\section*{Nomenclature}
\noindent The unit of mass [ma] can be [kg] or [mol].  Once set, it is to be used consistently.

\begin{center}
\begin{longtable}{|l l l|l l l|}
\hline
\multicolumn{3}{|c|}{Variables} & \multicolumn{3}{c|}{Porous medium} \\
$p$ & pressure & [Pa] & $\phi$ & porosity & [-] \\
$T$ & temperature & [K] & $\absperm$ & abs. permeability & [$\text{m}^2$] \\
$\specEnthalpy$ & specific enthalpy & [J / ma] & $k$ & rel. permeability & [-]\\
$z$ & overall component fraction & [-] & \multicolumn{3}{c|}{Mathematical notation} \\
$y$ & phase fraction & [-] & $\varX$ & flow \& transport variables & \\
$s$ & saturation & [-] & $\varY$ & equilibrium variables & \\
$x$ & physical partial fraction & [-] & $\sysF$ & system residual & \\
$\extfrac$ & extended partial fraction & [-] & $\jac$ & system Jacobian & \\

\multicolumn{3}{|c|}{Fluid properties} & $\alpha$ & increment step size & \\
$\rho$ & density & [ma / $\text{m}^3$] & \multicolumn{3}{c|}{Indexation} \\
$v$ & specific volume & [$\text{m}^3$ / ma] & $\ncomp$ & number of components & \\
$u$ & specific internal energy & [J / ma] & $\nphase$ & number of fluid phases & \\
$\mu$ & dynamic viscosity & [Pa s] & $_s$ & \multicolumn{2}{l|}{solid phase index (porous medium)} \\
$\kappa$ & thermal conductivity & [W / m K] & $_\xi$ & component index & \\
$\varphi$  & fugacity coefficient & [-] & $_\eta$ & \multicolumn{2}{l|}{contextual component index} \\

\multicolumn{3}{|c|}{Transport quantities} & $_\delta$ & phase index & \\
$\mathbf{v}$ & volumetric flux & [m / s] & $_\gamma$ & contextual phase index & \\
$\mathbf{q_e}$ & heat flux & [J m / s] & \multicolumn{3}{c|}{} \\
$\mathbf{m}$ & mass flux & [ma m / s] & \multicolumn{3}{c|}{} \\
$\lambda$ & mass mobility & [ma / s] & \multicolumn{3}{c|}{} \\
$\lambda_e$ & energy mobility & [J ma / s] & \multicolumn{3}{c|}{} \\

\hline
\end{longtable}
\end{center}
\FloatBarrier
\printbibliography

@article{voskov2017,
	title = {Operator-based linearization approach for modeling of multiphase multi-component flow in porous media},
	author = {{Voskov}, Denis V.},
	journal = {Journal of Computational Physics},
	year = {2017},
	volume = {337},
	pages = {275-288},
	doi = {10.1016/j.jcp.2017.02.041}
}

@article{lipovac2024,
title = {Unified flash calculations with isenthalpic and isochoric constraints},
journal = {Fluid Phase Equilibria},
volume = {578},
pages = {113991},
year = {2024},
issn = {0378-3812},
doi = {10.1016/j.fluid.2023.113991},
author = {Lipovac, Veljko and Duran, Omar and Keilegavlen, Eirik and Radu, Florin Adrian and Berre, Inga},
}

@article{Faigle2015,
  title = {Multi-physics modeling of non-isothermal compositional flow on adaptive grids},
  journal = {Computer Methods in Applied Mechanics and Engineering},
  volume = {292},
  pages = {16-34},
  year = {2015},
  note = {Special Issue on Advances in Simulations of Subsurface Flow and Transport (Honoring Professor Mary F. Wheeler)},
  issn = {0045-7825},
  doi = {10.1016/j.cma.2014.11.030},
  author = {Faigle, Benjamin and Elfeel, Mohamed Ahmed and Helmig, Rainer and Becker, Beatrix and Flemisch, Bernd and Geiger, Sebastian},
}

@article{moortgat2011,
  author = {Moortgat, Joachim and Sun, Shuyu and Firoozabadi, Abbas},
  title = {Compositional modeling of three-phase flow with gravity using higher-order finite element methods},
  journal = {Water Resources Research},
  volume = {47},
  number = {5},
  pages = {},
  doi = {10.1029/2010WR009801},
  year = {2011}
}

@article{chen2000,
  author = {Chen, Zhangxin},
  title = {Formulations and Numerical Methods of the Black Oil Model in Porous Media},
  journal = {SIAM Journal on Numerical Analysis},
  volume = {38},
  number = {2},
  pages = {489-514},
  year = {2000},
  doi = {10.1137/S0036142999304263},
}

@book{lake1989,
  author = {Lake, Larry W.},
  title = {Enhanced Oil Recovery},
  publisher = {Prentice Hall},
  year = {1989},
  isbn = {978-0132816014},
  address = {Englewood Cliffs, NJ}
}

@article{gharbia2021,
	title = {{An analysis of the unified formulation for the equilibrium problem of compositional multiphase mixtures}},
	author = {{Ben Gharbia}, Ibtihel and Haddou, Mounir and Tran, Quang Huy and Vu, Duc Thach Son},
	journal = {{ESAIM: Mathematical Modelling and Numerical Analysis}},
	year = {2021},
	volume = {55},
	number = {6},
	pages = {2981-3016},
	doi = {10.1051/m2an/2021075}
}

@article{lauser2011,
	title = {A new approach for phase transitions in miscible multi-phase flow in porous media},
	author = {Lauser, Andreas and Hager, Corina and Helmig, Reiner and Wohlmuth, Barbara},
	journal = {Adv. Water Resour.},
	year = {2011},
	volume = {34},
	number = {8},
	pages = {957-966},
	doi = {10.1016/j.advwatres.2011.04.021}
}

@article{keilegavlen2020,
	title = {PorePy: An Open-Source Software for Simulation of Multiphysics Processes in Fractured Porous Media}, 
	author = {Keilegavlen, Eirik and Berge, Runar and Fumagalli, Alessio and Starnoni, Michele and Stefansson, Ivar and Varela, Jhabriel and Berre, Inga},
	journal = {Comput. Geosci.},
	year = {2021},
	volume = {25},
	pages = {243-265},
	doi = {10.1007/s10596-020-10002-5}
}

@article{michelsen1982-1,
	title = {The isothermal flash problem. Part I. Stability},
	author = {Michelsen, Michael L.},
	journal = {Fluid Phase Equilib.},
	year = {1982},
	volume = {9},
	pages = {1-19},
	doi = {10.1016/0378-3812(82)85001-2}
}

@article{michelsen1982-2,
	title = {The isothermal flash problem. Part II. Phase-split calculation},
	author = {Michelsen, Michael L.},
	journal = {Fluid Phase Equilib.},
	year = {1982},
	volume = {9},
	pages = {21-40},
	doi = {10.1016/0378-3812(82)85002-4}
}

@article{gupta1990,
	title = {Simultaneous multiphase isothermal/isenthalpic flash and stability calculations for reacting/non-reacting systems},
	author = {Gupta, Anup K. and Bishnoi, P.Raj and Kalogerakis, Nicolas},
	journal = {Gas Sep. Purif.},
	year = {1990},
	volume = {4},
	number = {4},
	pages = {215-222},
	doi = {10.1016/0950-4214(90)80045-M}
}

@article{vu2021,
	title = {A new approach for solving nonlinear algebraic systems with complementarity conditions. Application to compositional multiphase equilibrium problems},
	author = {Vu, Duc Thach Son Vu and {Ben Gharbia}, Ibtihel and Haddou, Mounir and Tran, Quang Huy},
	journal = {Mathematics and Computers in Simulation},
	year = {2021},
	volume = {190},
	pages = {1243-1274},
	doi = {10.1016/j.matcom.2021.07.015}
}

@article{zhu2016,
	title = {Multiphase isenthalpic flash integrated with stability analysis},
	author = {Zhu, Di and Okuno, Ryosuke},
	journal = {Fluid Phase Equilib.},
	year = {2016},
	volume = {423},
	number = {},
	pages = {203--219},
	doi = {10.1016/j.fluid.2016.04.005},
}

@article{michelsen1999,
    title = {State function based flash specifications},
    author = {Michelsen, Michael L.},
    journal = {Fluid Phase Equilib.},
    year = {1999},
    volume = {158},   
    pages = {617--626},
    doi = {10.1016/S0378-3812(99)00092-8},
}

@article{nichita2024,
    title = {A unified presentation of phase stability analysis including all major specifications},
    journal = {Fluid Phase Equilibria},
    volume = {578},
    pages = {113990},
    year = {2024},
    issn = {0378-3812},
    doi = {10.1016/j.fluid.2023.113990},
    author = {Nichita, Dan Vladimir},
}

@article{zhu2014,
	title = {A robust algorithm for isenthalpic flash of narrow-boiling fluids},
	author = {Zhu, Di and Okuno, Ryosuke},
	journal = {Fluid Phase Equilibria},
	year = {2014},
	volume = {379},
	number = {},
	pages = {26-51},
	doi = {10.1016/j.fluid.2014.07.003},
}

@article{aavatsmark2002,
	title = {An {Introduction} to {Multipoint} {Flux} {Approximations} for {Quadrilateral} {Grids}},
	volume = {6},
	issn = {1573-1499},
	doi = {10.1023/A:1021291114475},
	number = {3},
	urldate = {2025-07-18},
	journal = {Computational Geosciences},
	author = {Aavatsmark, Ivar},
	month = sep,
	year = {2002},
	pages = {405--432},
}

@article{weiss2014,
	title = {Hydrothermal, multiphase convection of {H}$_{\textrm{2}}${O}‐{NaCl} fluids from ambient to magmatic temperatures: a new numerical scheme and benchmarks for code comparison},
	volume = {14},
	issn = {1468-8115, 1468-8123},
	shorttitle = {Hydrothermal, multiphase convection of {H}$_{\textrm{2}}${O}‐{NaCl} fluids from ambient to magmatic temperatures},
	doi = {10.1111/gfl.12080},
	language = {en},
	number = {3},
	journal = {Geofluids},
	author = {Weis, P. and Driesner, T. and Coumou, D. and Geiger, S.},
	month = aug,
	year = {2014},
	note = {Publisher: Wiley},
	pages = {347--371},
}

@article{trangenstein1989,
	title = {Mathematical {Structure} of {Compositional} {Reservoir} {Simulation}},
	volume = {10},
	issn = {0196-5204, 2168-3417},
	doi = {10.1137/0910049},
	language = {en},
	number = {5},
	journal = {SIAM J. Sci. and Stat. Comput.},
	author = {Trangenstein, John A. and Bell, John B.},
	month = sep,
	year = {1989},
	note = {Publisher: Society for Industrial \& Applied Mathematics (SIAM)},
	pages = {817--845},
}

@article{courant1952,
	title = {On the solution of nonlinear hyperbolic differential equations by finite differences},
	volume = {5},
	issn = {1097-0312},
	doi = {10.1002/cpa.3160050303},
	language = {en},
	number = {3},
	journal = {Communications on Pure and Applied Mathematics},
	author = {Courant, Richard and Isaacson, Eugene and Rees, Mina},
	year = {1952},
	pages = {243--255},
}

@article{peng1976,
	title = {A {New} {Two}-{Constant} {Equation} of {State}},
	volume = {15},
	issn = {0196-4313, 1541-4833},
	doi = {10.1021/i160057a011},
	language = {en},
	number = {1},
	journal = {Ind. Eng. Chem. Fund.},
	author = {Peng, Ding-Yu and Robinson, Donald B.},
	month = feb,
	year = {1976},
	note = {Publisher: American Chemical Society (ACS)},
	pages = {59--64},
}

@article{driesner2007,
	title = {The system {H2O}–{NaCl}. {Part} {I}: {Correlation} formulae for phase relations in temperature–pressure–composition space from 0 to 1000°{C}, 0 to 5000bar, and 0 to 1 {XNaCl}},
	volume = {71},
	issn = {0016-7037},
	shorttitle = {The system {H2O}–{NaCl}. {Part} {I}},
	doi = {10.1016/j.gca.2006.01.033},
	language = {en},
	number = {20},
	journal = {Geochimica et Cosmochimica Acta},
	author = {Driesner, Thomas and Heinrich, Christoph A.},
	month = oct,
	year = {2007},
	note = {Publisher: Elsevier BV},
	pages = {4880--4901},
}

@article{iapws1997,
	title = {The {IAPWS} Industrial Formulation 1997 for the Thermodynamic Properties of Water and Steam },
	author = {Wagner, W.  and Cooper, J. R.  and Dittmann, A.  and Kijima, J.  and Kretzschmar, H.-J.  and Kruse, A.  and Maresˇ, R.  and Oguchi, K.  and Sato, H.  and Stoecker, I.  and Sˇifner, O.  and Takaishi, Y.  and Tanishita, I.  and Truebenbach, J.  and Willkommen, Th. },
	journal = {Journal of Engineering for Gas Turbines and Power},
	year = {2000},
	volume = {122},
	pages = {150-184},
	issn = {0742-4795},
	doi = {10.1115/1.483186}
}

@article{wang2020,
	title = {An efficient numerical simulator for geothermal simulation: {A} benchmark study},
	volume = {264},
	issn = {0306-2619},
	shorttitle = {An efficient numerical simulator for geothermal simulation},
	doi = {10.1016/j.apenergy.2020.114693},
	language = {en},
	urldate = {2025-07-16},
	journal = {Applied Energy},
	author = {Wang, Yang and Voskov, Denis and Khait, Mark and Bruhn, David},
	month = apr,
	year = {2020},
	note = {Publisher: Elsevier BV},
	pages = {114693},
}

@article{younis2009,
	title = {Adaptively {Localized} {Continuation}-{Newton} {Method}—{Nonlinear} {Solvers} {That} {Converge} {All} the {Time}},
	volume = {15},
	issn = {1086-055X},
	doi = {10.2118/119147-PA},
	number = {02},
	journal = {SPE Journal},
	author = {Younis, R.M.. M. and Tchelepi, H.A.. A. and Aziz, K..},
	month = dec,
	year = {2009},
	pages = {526--544},
}

@article{novikov2025,
	title = {A finite volume framework for the fully implicit thermal-hydro-mechanical-compositional modeling in subsurface applications},
	volume = {538},
	issn = {0021-9991},
	doi = {10.1016/j.jcp.2025.114152},
	journal = {Journal of Computational Physics},
	author = {Novikov, Aleksei and Saifullin, Ilshat and Hajibeygi, Hadi and Voskov, Denis},
	month = oct,
	year = {2025},
	pages = {114152},
}

@article{flemisch2011,
	series = {New {Computational} {Methods} and {Software} {Tools}},
	title = {{DuMux}: {DUNE} for multi-\{phase,   component,   scale,   physics,   …\} flow and transport in porous media},
	volume = {34},
	issn = {0309-1708},
	shorttitle = {{DuMux}},
	doi = {10.1016/j.advwatres.2011.03.007},
	number = {9},
	journal = {Advances in Water Resources},
	author = {Flemisch, B. and Darcis, M. and Erbertseder, K. and Faigle, B. and Lauser, A. and Mosthaf, K. and Müthing, S. and Nuske, P. and Tatomir, A. and Wolff, M. and Helmig, R.},
	month = sep,
	year = {2011},
	pages = {1102--1112},
}

@article{class2002,
	title = {Numerical simulation of non-isothermal multiphase multicomponent processes in porous media.: 1. {An} efficient solution technique},
	volume = {25},
	issn = {0309-1708},
	shorttitle = {Numerical simulation of non-isothermal multiphase multicomponent processes in porous media.},
	doi = {10.1016/S0309-1708(02)00014-3},
	number = {5},
	journal = {Advances in Water Resources},
	author = {Class, H. and Helmig, R. and Bastian, P.},
	month = may,
	year = {2002},
	pages = {533--550},
}

@article{roy2020,
	title = {A {Constrained} {Pressure}-{Temperature} {Residual} ({CPTR}) {Method} for {Non}-{Isothermal} {Multiphase} {Flow} in {Porous} {Media}},
	volume = {42},
	issn = {1064-8275},
	doi = {10.1137/19M1292023},
	number = {4},
	journal = {SIAM Journal on Scientific Computing},
	author = {Roy, Thomas and Jönsthövel, Tom B. and Lemon, Christopher and Wathen, Andrew J.},
	month = jan,
	year = {2020},
	pages = {B1014--B1040},
}

@inproceedings{ogontula2025,
  title        = {A Unified Compositional Flow Model for Simulating Multiphase High-Enthalpy Geothermal Reservoirs},
  author       = {Oguntola, Michael and Duran, Omar and Lipovac, Veljko and Keilegavlen, Eirik and Berre, Inga},
  year         = 2025,
  month        = {3},
  booktitle    = {Proceedings of the 50th Workshop on Geothermal Reservoir Engineering Stanford University},
  publisher    = {Stanford},
  series       = {Stanford Geothermal Workshop},
  url = {https://hdl.handle.net/11250/3183366},
}

@inproceedings{duran2025,
  title        = {Mixed-Dimensional Approach for Compositional Multiphase Flow in High-Enthalpy Fractured Geothermal Reservoirs},
  author       = {Duran, Omar and Lipovac, Veljko and Berre, Inga},
  year         = 2025,
  month        = {3},
  booktitle    = {Proceedings of the 50th Workshop on Geothermal Reservoir Engineering Stanford University},
  publisher    = {Stanford},
  series       = {Stanford Geothermal Workshop},
  url = {https://hdl.handle.net/11250/3185251},
}

@article{armijo1966,
	title = {Minimization of functions having {Lipschitz} continuous first partial derivatives},
	volume = {16},
	issn = {0030-8730},
	url = {https://msp.org/pjm/1966/16-1/p01.xhtml},
	number = {1},
	journal = {Pacific Journal of Mathematics},
	author = {Armijo, Larry},
	month = jan,
	year = {1966},
	note = {Publisher: Mathematical Sciences Publishers},
	pages = {1--3},
}

@book{michelsen2004,
	title={Thermodynamic Modelling: Fundamentals and Computational Aspects},
	author={Michelsen, Michael L. and Mollerup, Jørgen},
	publisher={Tie-Line Publications},
	year={2004},
	isbn = {8798996118}
}

@article{voskov2012,
  author = {Voskov, Denis V. and Tchelepi, Hamdi A.},
  title = {Comparison of Nonlinear Formulations for Two-Phase Multi-Component EOS Based Simulation},
  journal = {Journal of Petroleum Science and Engineering},
  volume = {82-83},
  pages = {101--111},
  year = {2012},
  doi = {10.1016/j.petrol.2011.10.012},
  publisher = {Elsevier},
}

@article{kala2020,
  author = {Kala, K. and Voskov, D.},
  title = {Element Balance Formulation in Reactive Compositional Flow and Transport with Parameterization Technique},
  journal = {Computational Geosciences},
  volume = {24},
  number = {3},
  pages = {609--624},
  year = {2020},
  doi = {10.1007/s10596-019-9828-y},
  publisher = {Springer}
}

@inproceedings{li2023,
	title = {Fluid {Property} {Model} for {Carbon} {Capture} and {Storage} by {Volume}-{Translated} {Peng}-{Robinson} {Equation} of {State} and {Lohrenz}-{Bray}-{Clark} {Viscosity} {Correlation}},
	doi = {10.2118/212584-MS},
    booktitle = {SPE Reservoir Characterisation and Simulation Conference and Exhibition},
	language = {en},
	publisher = {OnePetro},
	author = {Li, Zhidong and Wanat, Edward and Lun, Lisa and Hoyt, Jordan and Heider, Andrew and Leahy-Dios, Alana and Wattenbarger, Robert},
	month = jan,
	year = {2023},
}

@article{lohrenz1964,
	title = {Calculating {Viscosities} of {Reservoir} {Fluids} {From} {Their} {Compositions}},
	volume = {16},
	issn = {0149-2136},
	doi = {10.2118/915-PA},
	number = {10},
	journal = {Journal of Petroleum Technology},
	author = {Lohrenz, John and Bray, Bruce G. and Clark, Charles R.},
	month = oct,
	year = {1964},
	pages = {1171--1176},
}

@article{gharbia2019,
	title = {Study of compositional multiphase flow formulation using complementarity conditions},
	volume = {74},
	issn = {1294-4475, 1953-8189},
	doi = {10.2516/ogst/2019012},
	journal = {Oil \& Gas Science and Technology – Revue d’IFP Energies nouvelles},
	author = {{Ben Gharbia}, Ibtihel and Flauraud, Eric},
	year = {2019},
	pages = {43},
}

\end{document}